\documentclass[11pt,a4paper]{article}
\usepackage{jheppub}
\usepackage[T1]{fontenc} 
\usepackage{subcaption}

\usepackage[normalem]{ulem}
\usepackage{float}
\usepackage{caption}
\usepackage[english]{babel}
\usepackage{adjustbox}
\usepackage{pdflscape}
\usepackage{afterpage}
\usepackage{capt-of}

\usepackage{lipsum}

\newcommand{\br}{\begin{eqnarray}}
\newcommand{\er}{\end{eqnarray}}

\def\invfb{\text{fb}^{-1}}

\def\t {\widetilde {t_1}}

\def\C1{\widetilde \chi_1^{\pm}}
\def\mst1 {m_{\t1}}
\def\br {\begin{eqnarray}}
\def\er {\end{eqnarray}}

\def\invfb{\text{fb}^{-1}}

\def \Otphi{\mathcal{O}_{t\phi}}

\def \Opqthree{\mathcal{O}_{\phi Q}^{(3)}}

\def \Opq1{\mathcal{O}_{\phi Q}^{(1)}}
\def \Otw{\mathcal{O}_{tW}}
\def \Opw{\mathcal{O}_{\phi W}}
\def \Ctphi{C_{ t\phi}}

\def \Cpq1{C_{\phi Q}^{(1)}}
\def \Cpq3{C_{\phi Q}^{(3)}}

\def \Ctw{C_{tW}}
\def \Cpw{C_{\phi W}}
\def \Cqq31{C_{Qq}^{(3,1)}}
\def \Cqq38{C_{Qq}^{(3,8)}}
\mathchardef\mhyphen="2D 

\title{ Exploring SMEFT operators in the tHq production at the LHC}
\author{\small Monoranjan Guchait~}
\author{\small and ~Arnab Roy}
\affiliation{\small Department of High Energy Physics,
	Tata Institute of Fundamental Research,\\~~~~~~
	Homi Bhabha Road, Mumbai-400005, India}
\emailAdd{guchait@tifr.res.in}
\emailAdd{arnab.roy@tifr.res.in}

\abstract{
		We study the top-quark production along with a Higgs boson and a jet (tHq) at the LHC experiment within the framework of the Standard Model Effective Field Theory (SMEFT). A strategy is developed to constrain the Wilson Coefficients (WC) corresponding to the associated SMEFT operators using the latest LHC measurements. The best-fit values of these WCs are presented. Finally, we demonstrate the feasibility of finding the effects of these operators on various kinematical observables of the tHq process at the LHC. We find that for a set of best-fitted values of the considered WCs, the excess of signal over the backgrounds can be achieved with a reasonable significance at the center of mass energy $\sqrt{s}=13$ TeV and for integrated luminosity options ${\cal L}=$300~$\invfb$ and 3000~$\invfb$.}

\begin{document}
 		\maketitle
 	\flushbottom
\section{Introduction}
Several measurements along with precision tests carried out at LHC experiments, CERN, Geneva, have re-established the theory of interactions among the elementary particles, namely the
Standard Model (SM).
However, the shortcomings of the SM from the theoretical and as well experimental sides motivate us to think beyond the SM (BSM) physics. Recall, the SM is constructed based on the $\rm SU(3)_c\times SU(2)_L\times U(1)_Y$ gauge symmetry. Therefore, it is natural to expect that the proposed BSM models should be invariant under this gauge group at the electroweak (EW) scale, and respect all measurements that have been done so far in various experiments. Moreover, the lack of evidence of any kind of new physics signal at the current range of energy scale may suggest that the masses of the BSM particles reside at a higher scale or are decoupled from the SM. Intuitively, one can argue that the SM can be thought of as a low-energy manifestation of a more general fundamental description at a very high energy scale, say $\Lambda$~\cite{Appelquist:1974tg,Weinberg:1979sa,Weldon:1980gi,Eichten:1983hw,Brivio:2017vri}. In this approach, the SM effective theory (SMEFT) is regarded as a useful framework to describe such theories. In recent times, the SMEFT framework has received a lot of attention and has become a popular choice to study the effect of UV scale physics in a more generalized and comprehensive manner~\cite{Buchmuller:1985jz,Grzadkowski:2010es}.

In the SMEFT prescription, the SM lagrangian is expanded by adding
the additional higher order terms~\cite{Buchmuller:1985jz,Grzadkowski:2010es,article_rao},
which can be described as,  
\br
\mathrm{\mathcal{L}_{SMEFT}= \mathcal{L}_{SM}}+\mathrm{\frac{1}
{\Lambda^2}\sum_{i=1}^{N_{d6}}} C_i^{(6)}\mathcal{O}_i^{(6)}.
\label{eq:smeft}
\er
Here ${\cal L}_{\rm SM}$ corresponds to the usual SM lagrangian, and the second term represents the EFT part consisting of total $\rm N_{d6}$ number of dimension-6 operators ${\cal O}_i^{(6)}$,
constructed out of SM fields with coefficients $C_i^{(6)}$ and cut-off scale $\rm \Lambda$, preserving $\rm SU(3)_c\times SU(2)_L\times U(1)_Y$ gauge invariance. Here $\Lambda$ is a dimensionful parameter and $C_i$'s, which are dimensionless, are known as the Wilson coefficients (WC). In principle, there exist a total of 2499 dimension-6 operators assuming lepton
and baryon number conservation, and once the flavor universality is imposed, this number
comes down to 59~\cite{Grzadkowski:2010es}. Notably, there can be only one SMEFT operator with dimension-5 in this set-up, known as the Weinberg operator, which generates the lepton number violating Majorana mass term for left-handed neutrinos. However, this operator is not relevant to the physics we address in this study.

Obviously, the addition of extra dimension-6 terms in the lagrangian 
(Eq.~\ref{eq:smeft}) results in the modification of interactions 
from its standard form predicted by the SM.  
Consequently, the contribution due to the SMEFT operators
may cause deviations in SM predictions for precision measurements,  
which in turn constrain the new physics effects  
and may impose restrictions on the corresponding WCs. 
It is found that a class of SMEFT operators affects a large   
variety of known SM processes in a moderate to severe scale~\cite{BessidskaiaBylund:2016jvp,Dawson:2018dxp,Hu:2018veh,DHondt:2018cww,Degrande:2018fog,Dawson:2020oco,Goldouzian:2020wdq,Aebischer:2021uvt,Ethier:2021ydt,Khatibi:2020mvt,Araz:2020zyh,Boughezal:2020klp,Boughezal:2021tih,Battaglia:2021nys,Bellan:2021dcy,Afik:2021xmi,Faham:2021zet,Dawson:2021xei,Ellis:2021dfa,Allwicher:2022gkm,Boughezal:2022nof,Barman:2022vjd,Haisch:2022nwz,Kim:2022amu,Goldouzian:2020ekx}.
Currently, several studies, as reported in the literature, 
have phenomenologically constrained a larger set of SMEFT operators 
fitting a combined set of numerous 
measurements~\cite{Han:2004az,Corbett:2012ja,Chang:2013cia,Corbett:2013pja,Dumont:2013wma,Pomarol:2013zra,Elias-Miro:2013mua,Boos:2013mqa,Ellis:2014dva,Ellis:2014jta,Falkowski:2014tna,Efrati:2015eaa,Falkowski:2015krw,Buckley:2015lku,Berthier:2016tkq,Falkowski:2017pss,Ellis:2018gqa,Banerjee:2018bio,daSilvaAlmeida:2018iqo,Aebischer:2018iyb,Biekoetter:2018ypq,Descotes-Genon:2018foz,Silvestrini:2018dos,Hartland:2019bjb,Hurth:2019ula,vanBeek:2019evb,Durieux:2019rbz,Bissmann:2019gfc,Brivio:2019ius,Falkowski:2019hvp,Banerjee:2019twi,CMS:2020pnn,ATLAS:2020naq,CMS:2020gsy,Aoude:2020dwv,Biekotter:2020flu,Faroughy:2020ina,Aebischer:2020lsx,Aebischer:2020dsw,Falkowski:2020pma,Falkowski:2020znk,Ellis:2020unq,Ethier:2021bye}
carried out at the LHC experiments.
Dedicated searches of new physics effects due to SMEFT operators 
are also carried out widely by the ATLAS and CMS collaborations, 
and from the non-observation of any excess of events, the excluded 
ranges of concerned SMEFT operators are 
predicted~\cite{ATLAS:2018zsq,ATLAS:2019mke,ATLAS:2019fwo,ATLAS:2020rej,
	ATLAS:2021atp,ATLAS:2021kog,ATLAS:2021dqb,ATLAS:2022vym,
	ATLAS:2022xfj,ATLAS:2021stq,ATLAS:2021zem,ATLAS:2022mlu,
	ATLAS:2022tnm,CMS:2019zct,CMS:2019nrx,CMS:2019jsc,CMS:2019too,
	CMS:2020lrr,CMS:2021nnc,CMS:2021aly,CMS:2021mmh,CMS:2021yzl}. 

In this study, more emphasis is given to explore the feasibility of finding the signature of SMEFT operators at the LHC. As we understand, the inclusion of the SMEFT operators in the SM lagrangian, which is very well tested, may lead to a change in the total as well as in the differential cross-sections. In particular, the interference effects 
between amplitudes of the SM, and the SMEFT operators may show up very likely at the tail in the differential distributions of cross-sections, of course, depending on the sensitivity of the corresponding set of operators to the given process. Looking for signals of all SMEFT operators together is undoubtedly very complicated, and even may not be feasible at the LHC. Instead, in this study, we consider a different avenue focusing only on a subset of SMEFT operators related to a particular interesting physics process. In this approach, we found that the top quark production in association with a Higgs boson
and a jet,
\br
\rm pp \to tHq,
\label{eq:thq}
\er
 is one such interesting choice. This process includes the Yukawa coupling t-t-H which has several significances. For instance, it is the strongest one and a potential source to explore the effects of new physics and the EWSB~\cite{Bezrukov:2014ina,CARENA1994213,Delepine:1995qs,Chivukula:1998wd}. Already, various precision measurements are carried out at the LHC to test this coupling~\cite{Chang:2014rfa,CMS:2015nrd,CMS:2017con,
	CMS:2018hnq,CMS:2020cga,CMS:2020mpn,ATLAS:2014ayi,ATLAS:2015utn, ATLAS:2016wki,ATLAS:2018mme,ATLAS:2021qou,ATLAS:2022yrq,ATLAS:2022tnm}. 
In addition, the presence of the W-t-b vertex in the above process 
is also another aspect where the impact of SMEFT operators is known to be significant. The W-t-b vertex can bear the imprints of different UV models~\cite{Dabelstein:1995jt,Cao:2003yk,delAguila:2000rc,Belyaev:2006jh,Contino:2006nn,Aguilar-Saavedra:2009xmz,Cao:2012ng} through the anomalous coupling. 
Eventually, the combined effects of the SMEFT operators to 
the t-t-H and W-t-b vertices may lead to some interesting phenomena. For instance, some of the relevant operators lead to energy-growing features in scattering amplitudes~\cite{Farina:2012xp}, resulting in excess in the kinematic distributions of some observables, as we will see at the end. Hence, the prospect of the tHq process in probing this kind of high-scale physics through the SMEFT framework seems to be very promising.

Our study is performed in two steps. Firstly, the set of sensitive SMEFT operators affecting the tHq process is identified. It is realized that the same set of SMEFT operators can also affect some other production processes at the LHC involving top quark in the final state.   
The deviations of various measurements, such as total
and differential cross-sections, signal strengths, asymmetries, etc. 
from the corresponding theoretical predictions 
of those processes are used to 
impose constraints on the associated SMEFT operators. 
Notably, these SMEFT operators are constrained from global fits also. However, we revisit those constraints only focusing on the most recent measurements sensitive to the chosen set of operators.
 Secondly, the implications of those constrained operators to various 
kinematic distributions of the tHq process are 
investigated. Eventually, we demonstrate the discovery potential of the signatures of SMEFT operators presenting the signal significance observed at the tail of the distributions. We present results for the high luminosity options, $\rm \mathcal{L}=300~fb^{-1}$ and $\rm 3000~fb^{-1}$, for the LHC experiment with centre of mass energy $\rm \sqrt{s}=13~TeV$. The SMEFT effects on the basic 
kinematic distributions of the tHq process are studied 
before~\cite{Degrande:2018fog}, but a detailed picture with proper 
detector simulation, and sensitivity projections at future 
LHC luminosity options, are found to be absent. 
Thus, revisiting the tHq production in the SMEFT context, 
this study points out the importance of this process to single out 
EFT effects, with detailed simulation and analysis.  

This paper is organized as follows. 
We first identify the number of sensitive SMEFT operators which modify the 
vertices of the tHq process in section 2. In section 3, considering 
various measurements at the LHC relevant to 
these operators, a methodology is developed by which we predict the 
best-fit values of the corresponding WCs of those operators interesting to us, 
along with their allowed ranges. Cross 
sections of the relevant processes are calculated in section 3, including 
a formulation to construct Fisher information matrix to correlate the
sensitivities of SMEFT operators with them. The effects of those operators in 
various kinematic distributions of the final state particles of the 
tHq process are studied in section 4. Finally, in the same section, 
estimating the contribution of various backgrounds in the signal excess region, 
we present the signal significances in several kinematic bins.   
A summary with concluding remarks is presented in section 5.

\section{SMEFT operators for tHq}
\begin{figure}[H]
	\begin{subfigure}[b]{0.3\textwidth}
		\centering
		\includegraphics[width=3.0 cm]{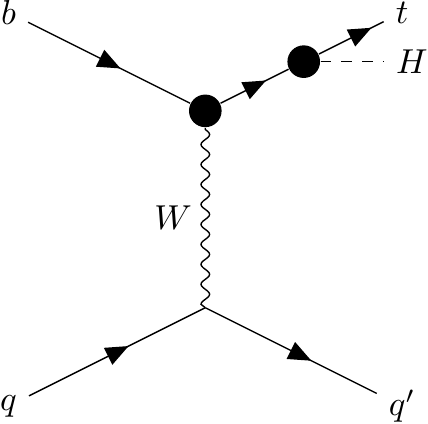}
	\end{subfigure}
	\begin{subfigure}[b]{0.3\textwidth}
		\centering
		\includegraphics[width=3.0 cm]{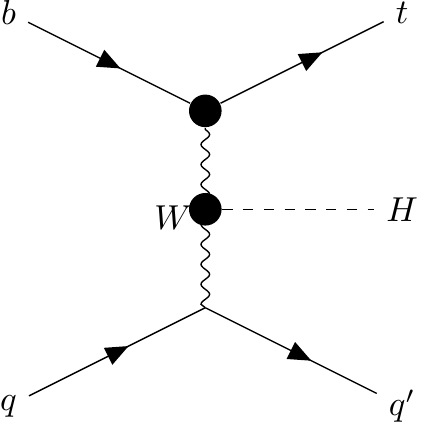}
	\end{subfigure}
	\begin{subfigure}[b]{0.3\textwidth}
		\centering
		\includegraphics[width=3.0 cm]{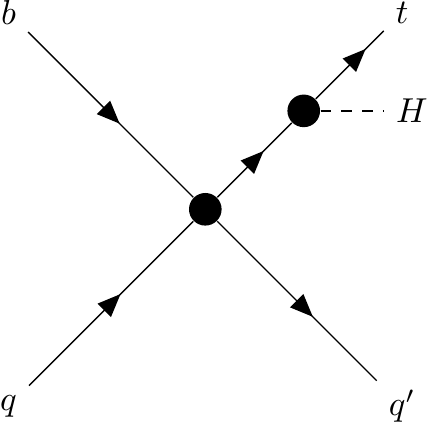}
	\end{subfigure}
	\caption{\small {Feynman diagrams of the tHq process in the t-channel.}}
	\label{fig:feyndiag}
\end{figure}

In this section we identify the set of SMEFT operators associated with the process Eq.~\ref{eq:thq}. The representative Feynman diagrams at the tree level for the above process are shown in Fig.~\ref{fig:feyndiag}, where the blobs represent the vertices modified by the SMEFT operators.

In this study, we explore the effect of EFT to the t-t-H coupling as well as to the top quark chirality-flipping (W-t-b) interactions through the tHq process. 
On the other hand, following the standard practice, we 
neglect the effects of EFTs to the other vertices not involving top-quark~\footnote{Effect of EFT to Wqq vertices are expected to be negligible due to strong constraints on the relevant operators from electroweak precision observables (EWPO)~\cite{ALEPH:2005ab}.}.
It can be imposed by invoking  the symmetry requirement as,
\br
U(3)_l\times U(3)_e\times U(2)_Q\times U(2)_u \times U(3)_d\equiv U(2)^2 \times U(3)^3,
\label{eq:flavor_sym}
\er
closely following the prescription of the LHC Top Quark Working Group~\cite{Aguilar-Saavedra:2018ksv}, guided by Minimal Flavour 
Violation (MFV) hypothesis~\cite{DAmbrosio:2002vsn}. 
This assumption is also adopted in the calculations implemented in \nolinkurl{SMEFTatNLO}~\cite{Degrande:2020evl}, which we use to compute the cross-sections including EFT effects.

Under this flavor assumption, a larger set of SMEFT operators reduces to seven most relevant operators for the tHq process as shown in Table~\ref{tab:op_prod}.  Here, the Higgs doublet is denoted by $\phi$, while $\tilde{\phi}\equiv i\tau^2\phi$; Q and t are the left-handed quark doublet and right-handed singlet respectively; $W^I_{\mu\nu}$ is the $SU(2)_L$ weak gauge field strength tensor and $\rm D^\mu$ is the covariant derivative, with $(\phi^{\dagger}\overleftrightarrow{D_{\mu}}\phi)\equiv \phi^{\dagger}({D_{\mu}}\phi)-(\phi^{\dagger}{D_{\mu}})\phi$. These set of operators modify the couplings, such as W-t-b, t-t-H, and W-W-H involved in the tHq production process as shown in Fig.~\ref{fig:feyndiag}.  
The point to note here that, these set of operators not only affect the tHq process, but several other important processes involving top-quark and Higgs boson, as shown in the last column of Table~\ref{tab:op_prod}, are also equally affected. All of these processes are studied at the LHC RUN2 experiments by measuring the corresponding cross-sections and other related observables~\cite{CMS:2018jeh,CMS:2020mpn,CMS:2021ugl,ATLAS:2022vym,CMS:2021ugv,CMS:2021vqm,ATLAS:2021fzm}.
Naturally, any deviation of these measurements from the corresponding SM theoretical predictions including EFT, probably can constrain the WCs of these set of operators. With this intuition, we attempt to find the allowed ranges of the corresponding WCs, consistent with experimental measurements, and eventually predict their best fit values.
\begin{table}[H]
	\caption{\small {SMEFT operators contributing to the $\rm tHq$ production (Fig.\ref{fig:feyndiag}) in the Warsaw basis, with an assumption (Eq.~\ref{eq:flavor_sym})~\cite{Aguilar-Saavedra:2018ksv}.}}
	\centering 
	\begin{tabular}{c c c c c}	
		\hline 
		\hline
		Operator & Coefficient   & Definition & Coupling & Sensitive processes \\
		\hline
		
		$\mathcal{O}_{ t\phi}$ & $C_{t\phi}$ & $(\phi^{\dagger}\phi-v^2/2)\bar{Q}t\tilde{\phi}$& $\rm t\mhyphen\bar{t}\mhyphen H$ & $\rm t\bar{t}H$, tHq \\
		
		$\mathcal{O}_{\phi Q}^{(3)}$ &$C_{\phi Q}^{(3)}$ &$i(\phi^{\dagger}\overleftrightarrow{D_{\mu}}\tau_{I}\phi)\bar{Q}\gamma^{\mu}\tau^IQ$ & W-t-b & tHq, tj, tV, $\rm t\bar{t}H$\\
		
		$\mathcal{O}_{tW}$ & $C_{tW}$ & $i(\bar{Q}\sigma^{\mu\nu}\tau_It)\tilde{\phi}W^I_{\mu\nu}$& W-t-b &  tHq, tj, tV, $\rm t\bar{t}H$\\
		$\mathcal{O}_{Qq}^{(3,1)}$ & $C_{Qq}^{(3,1)}$ & $ (\bar{q}_i\gamma^\mu\tau_I q_i)(\bar{Q}\gamma^\mu \tau^IQ)$& 4-Fermion &  tHq,tj,$\rm t\bar{t}$, $\rm t\bar{t}H$, $\rm t\bar{t}V$\\
		$\mathcal{O}_{Qq}^{(3,8)}$ & $C_{Qq}^{(3,8)}$ & $ (\bar{q}_i\gamma^\mu\tau_IT_A q_i)(\bar{Q}\gamma^\mu\tau^IT^AQ)$& 4-Fermion &  tHq,tj,$\rm t\bar{t}$, $\rm t\bar{t}H$, $\rm t\bar{t}V$\\
		$\mathcal{O}_{\phi W}$ & $C_{\phi W}$ & $(\phi^{\dagger}\phi-v^2/2)W_{\mu\nu}^{I} W_I^{\mu\nu}$& H-W-W &   tHq, $\rm t\bar{t}H$\\
		$\mathcal{O}_{\phi D}$ & $C_{\phi D}$ & $(\phi^{\dagger}{D_{\mu}}\phi)(\phi^{\dagger}{D^{\mu}}\phi)$& H-W-W &  tHq, $\rm t\bar{t}H$\\
		
		\hline
		\hline 
	\end{tabular}
	
	\label{tab:op_prod}
\end{table} 	
Notice that, among these seven operators, the  $\mathcal{O}_{\phi W}$ and $\mathcal{O}_{\phi D}$ mainly affect the Higgs boson coupling with W boson.  $\mathcal{O}_{\phi D}$ is expected to be 
tightly constrained by the measurements of the 
electroweak precision observables (EWPO)~\cite{ALEPH:2005ab}.  
Therefore, the effects of $\mathcal{O}_{\phi D}$ is not taken into account in our analysis.

The impacts of some of the other relevant SMEFT operators on the decays of 
the final state particles in the tHq process, in particular, 
$\rm H\to b\bar{b},W^+W^-,ZZ,\tau^+\tau^-,\gamma\gamma$, and $\rm t\to b W$, 
are also checked. The WC ($C_{b\phi}$) corresponding to the operator $\mathcal{O}_{b\phi}\equiv(\phi^{\dagger}\phi)\bar{Q}b\phi$ (though violates the symmetry requirement, Eq.~\ref{eq:flavor_sym}), modifying the $\rm H\mhyphen b\mhyphen \bar{b}$ coupling, may affect the width of our considered decay channel $\rm H\to b\bar{b}$.
However, it is found that $C_{b\phi}$ is severely 
constrained already from several
studies~\cite{Ethier:2021bye,Ellis:2020unq,ATLAS:2021vrm,ATLAS:2022xyx},
and within its constrained range ($\sim \pm 10^{-3}$), 
the basic kinematic distributions of the tHq process do not receive any visible effect. 
Thus, we have not included the impact of $\mathcal{O}_{b\phi}$ on the $\rm H\to b \bar b$ decays
in our analysis. 
Moreover, the impact of other SMEFT operators to the
H-V-V(V=W,Z) couplings, such as  $\mathcal{O}_{\phi W/D}$ may have some effect to the total decay width of Higgs. However, due to the strong constraints on $\mathcal{O}_{\phi D}$, as argued earlier, we do not consider this operator in Higgs boson decays and move on with $\mathcal{O}_{\phi W}$ only.
On the contrary, EFT effects on the top quark decay vertex (W-t-b) are taken into consideration as it is expected to provide a similar impact as in the production vertex. Thus the $\mathcal{O}_{\phi Q}^{(3)}$, and $\mathcal{O}_{tW}$ play important roles through the W-t-b vertex in the tHq production and subsequent decay of the top-quark. The other operator, such as $\Otphi$ affects the tHq production through the $\rm t\mhyphen\bar{t}\mhyphen H$ vertex, whereas $\mathcal{O}_{Qq}^{(3,1)}$ and $\mathcal{O}_{Qq}^{(3,8)}$ contribute via the 4-Fermion (4-F) interaction, including additional Feynman diagrams.
 
Finally, having all these considerations, we end up with the following six relevant set of operators for our present study,
\br
\mathcal{O}_{ t\phi},\mathcal{O}_{\phi Q}^{(3)},\mathcal{O}_{tW},\mathcal{O}_{Qq}^{(3,1)},\mathcal{O}_{Qq}^{(3,8)}, \Opw.
\label{eq:operators}
\er 
Now onward, in the text, SMEFT operators will refer to this set of six operators only. It is to be noted that several studies based on global fits 
have already restricted the WCs of these operators, 
considering a set of large number of experimental measurements~\cite{Ethier:2021bye,Ellis:2020unq}. 
In this present study, we revisit the constraints of 
the WCs corresponding to these set of SMEFT operators (Eq.~\ref{eq:operators}) taking into account the latest experimental measurements, which are sensitive 
to these operators. In addition, some of the new measurements 
from CMS and 
ATLAS experiments, that were not taken into consideration in the 
previous studies~\cite{ATLAS:2018mme,CMS:2018hnq,CMS:2021ugl,CMS:2018fdh,CMS:2018jeh,ATLAS:2021fzm,CMS:2021ugv,CMS:2021vqm,CMS:2022hjj}, are included in our analysis. In the next section, we systematically describe our methodology of constraining these operators.

\section{Constraining operators $\mathcal{O}_{ t\phi},\mathcal{O}_{\phi Q}^{(3)},\mathcal{O}_{tW},\mathcal{O}_{Qq}^{(3,1)},\mathcal{O}_{Qq}^{(3,8)},\Opw$}
Setting the goal to constrain the set of chosen SMEFT operators, first we set up the framework to calculate the theoretical cross-section and other related observables of several physics processes of our interest in terms of the WCs. Subsequently, in order to understand the sensitivity of the SMEFT operators of our interest to any given physics process, we construct the Fisher Information Matrix (FIM)~\cite{Brehmer:2016nyr}. Finally, performing a $\chi^2$-fitting of the calculated expressions of the observables with the corresponding experimental measurements, best-fit values of the WCs along with the constrained ranges are presented.
{\subsection{Cross-section calculation}}

Inclusion of SMEFT operators to the SM lagrangian (Eq.~\ref{eq:smeft}) eventually leads to the modification of cross-sections and various related observables. 
Restricting terms only up to dimension-6 operators in the EFT expansion,
the scattering amplitude of any physical process is given by,
\br
\mathcal{M}_{tot}= \mathcal{M}_{SM}+\mathcal{M}_{EFT}(C_i),
\er
and hence, the matrix element can be written as,
\br
|\mathcal{M}_{tot}(C_i)|^2=|\mathcal{M}_{SM}|^2
+2Re[\mathcal{M}_{SM}^*\mathcal{M}_{EFT}(C_i)]+|\mathcal{M}_{EFT}(C_i)|^2.
\label{eq:amp}
\er
Here $|\mathcal{M}_{SM}|^2$ resembles the SM part, while
the second term corresponds to the interference between EFT and SM
($\propto 1/\Lambda^2$), which is generally responsible for the dominant EFT contribution. The third term being the quadratic one, is suppressed by
$\propto 1/\Lambda^4$. The $C_i$ represents the WC for the ith operator, $\mathcal{O}_i$. Finally, the cross-section can be expressed in a more compact form as,
\br
\sigma^{EFT}({C})=\sigma^{SM}+\sum_{i=1}^{n} C_i\beta_i+\sum_{j\leq k}^{n}C_jC_k\gamma_{jk}.
\label{eq:sigma}
\er
Here $\sigma^{SM}$ is the standard model
cross-section, `${\rm n}$' is the total number of
relevant SMEFT operators taken into account to the given process, while $\rm \beta,\gamma$ are
process-dependent coefficients for the
linear and quadratic terms respectively, setting $\rm \Lambda=1~TeV$. From now onwards, all numerical calculations in this study will assume $\rm \Lambda=1~TeV$. In general, in the case of differential cross-section calculation, these coefficients depend on the momenta and masses of particles involved in the process. While integrating out over momenta to obtain the total cross-section, these remain mass dependent only. Needless to say, the SM cross-section can be recovered by setting all the $C_i$ to zero.

For some specific cases, the experimental measurements
are also presented in terms of signal strengths, differential cross-section,
asymmetry, etc., in addition to the total
production cross-sections. Interestingly, all of these observables
can be expressed, in a similar manner, in terms of a polynomial as Eq.~\ref{eq:sigma}.
For instance, the signal
strength for a given process,
say, $\rm p p \to tHq$ with t/H$\to xx$, where $xx$ symbolizes the decay products, one can write,
\br
\mu^{EFT}_{xx}({C})=\frac{\sigma^{EFT}_{xx}}{\sigma^{SM}_{xx}}=
1+\sum_{i=1}^{n}a_i C_i+\sum_{j\leq k}^{n}b_{jk}C_jC_k.
\label{eq:mu}
\er
Here, the coefficients $a_{i}$ and $b_{jk}$ are
the function of momenta and masses of particles involved in the process and specific to particular decay channel t/H$\to xx$. The symbol
``$xx$'' being removed from the polynomial expression for simplicity (see Appendix A).
The Eq.~\ref{eq:mu} is the most general expression for $\rm \mu^{EFT}_{xx}$ up to the quadratic 
order in WCs for the full process including the decay chain
(see Appendix A). Following the similar procedure, any observable related to the cross-section
can be expressed as a polynomial in WCs having
some process-dependent coefficients, such as, $\beta_i,\gamma_{jk}$ or $a_i, b_{jk}$ and so on.

In principle, the lagrangian including the SMEFT terms (Eq.~\ref{eq:smeft}) provides a framework which is calculable and predictive to find the impact of those operators to any process. However, it is quite complicated and also very exhaustive to obtain the analytical expressions of these coefficients, such as $\beta_i, \gamma_{ij}$ or $a_i, b_{ij}$ etc. for any chosen process. 
Instead, we follow a different approach. The numerical values of these coefficients are obtained by fitting of the cross-sections considering WCs as variables while $\beta_i, \gamma_{ij}$ as fitting parameters. In this approach, first the variation of cross-section with the WCs is obtained, which is then fitted with an appropriate polynomial to find the numerical values of the corresponding coefficients. This strategy is described in detail below.

The cross-sections are calculated at the leading order (LO) approximation using \nolinkurl{MG5aMC_atNLO}~\cite{Alwall:2014hca} interfacing with \nolinkurl{SMEFTatNLO}~\cite{Degrande:2020evl} feynrules UFO. While doing the calculations, we employ five-flavour scheme (5 FS) with \nolinkurl{NNPDF23LO} \cite{Ball:2010de} for parton distribution function and set the dynamic QCD scale ($Q^2=\frac{1}{2}\sum_{i=1}^{N_f}(m_i^2+p_{Ti}^2)$; $N_f$= number of final state particles).
A basic set of event generation kinematic selection cuts are applied
wherever required, to obtain cross-sections only in the
relevant fiducial region of phase space of any given process. The decays, wherever necessary, were performed using \nolinkurl{Madspin}, allowing us to take into account the spin of mother particle into decay products.

\begin{figure}
	\centering
	\includegraphics[width=7.5 cm]{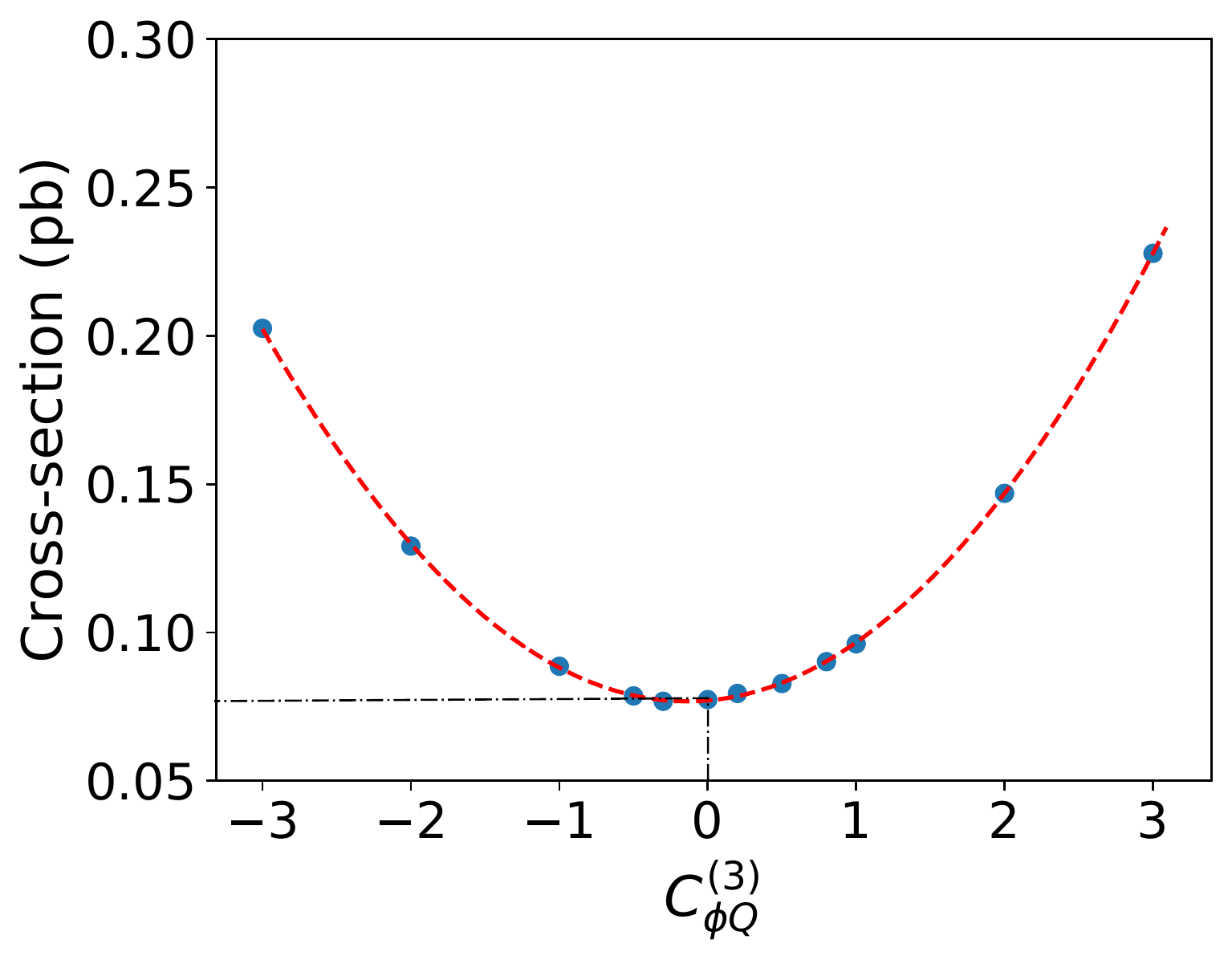}
	\caption{\small {The cross-section of tH production process for different values of $\rm C_{\phi Q}^{(3)}$ (solid points). The cross-section is fitted (dashed line) with a quadratic polynomial (Eq.~\ref{eq:thq_cpq3}).}}
	\label{fig:pq3}
\end{figure}

Before presenting the full calculation, first we demonstrate our strategy by giving an example of calculating the cross-section of a process including a single WC (say $\rm C_1$), and setting others to zero. With this assumption, any production cross-section (Eq.~\ref{eq:sigma}), with $\rm t/H\to xx$, can be written as,
\br
\rm \sigma^{EFT}_{xx}(C_1)=\sigma^{SM}+\beta_1C_1+\gamma_{11}C_1^2.
\label{eq:fit_quad}
\er
For example, the variation of $\rm \sigma^{EFT}_{xx}(C_1)$ for the tH (tHq+tHW) production for a range of values of $\rm C_1\equiv C_{\phi Q}^{(3)}$ produces a curve, quadratic in nature (Eq.~\ref{eq:fit_quad}), as shown in
Fig.~\ref{fig:pq3} by solid points.

Notice that the variation of the cross-section with WCs is almost symmetric around
$\rm C_{\phi Q}^{(3)}=0$, i.e, does not depend on the sign of WC. It implies
that the coefficient ($\rm \beta_1$) of the linear term
in Eq.~\ref{eq:fit_quad} is insignificant in comparison to
that of the quadratic term ($\rm \gamma_{11}$).
Now we perform a polynomial fitting of these points representing the cross-section. The fitted curve is shown as a dashed line in Fig~\ref{fig:pq3}, and the corresponding polynomial expression is given by,

\br
\rm \sigma^{EFT}_{xx}(\Cpq3) [pb]=0.077+0.004\:\Cpq3+0.015\: {\Cpq3}^2.
\label{eq:thq_cpq3}
\er
It confirms our observation that the interference term (with coefficient $\beta_1=0.004$) contributes to negligible level in this case. Needless to say that the cross-section for $C_{\phi Q}^{(3)}=C_1=0$ represents for SM tH production cross-section, $\sigma^{SM}=0.077$ pb,
and it is found to be consistent with the existing calculation in the literature~\cite{LHCHiggsCrossSectionWorkingGroup:2016ypw}.

\begin{figure}
	\begin{subfigure}[b]{0.5\textwidth}
		\centering
		\includegraphics[width=7.5 cm]{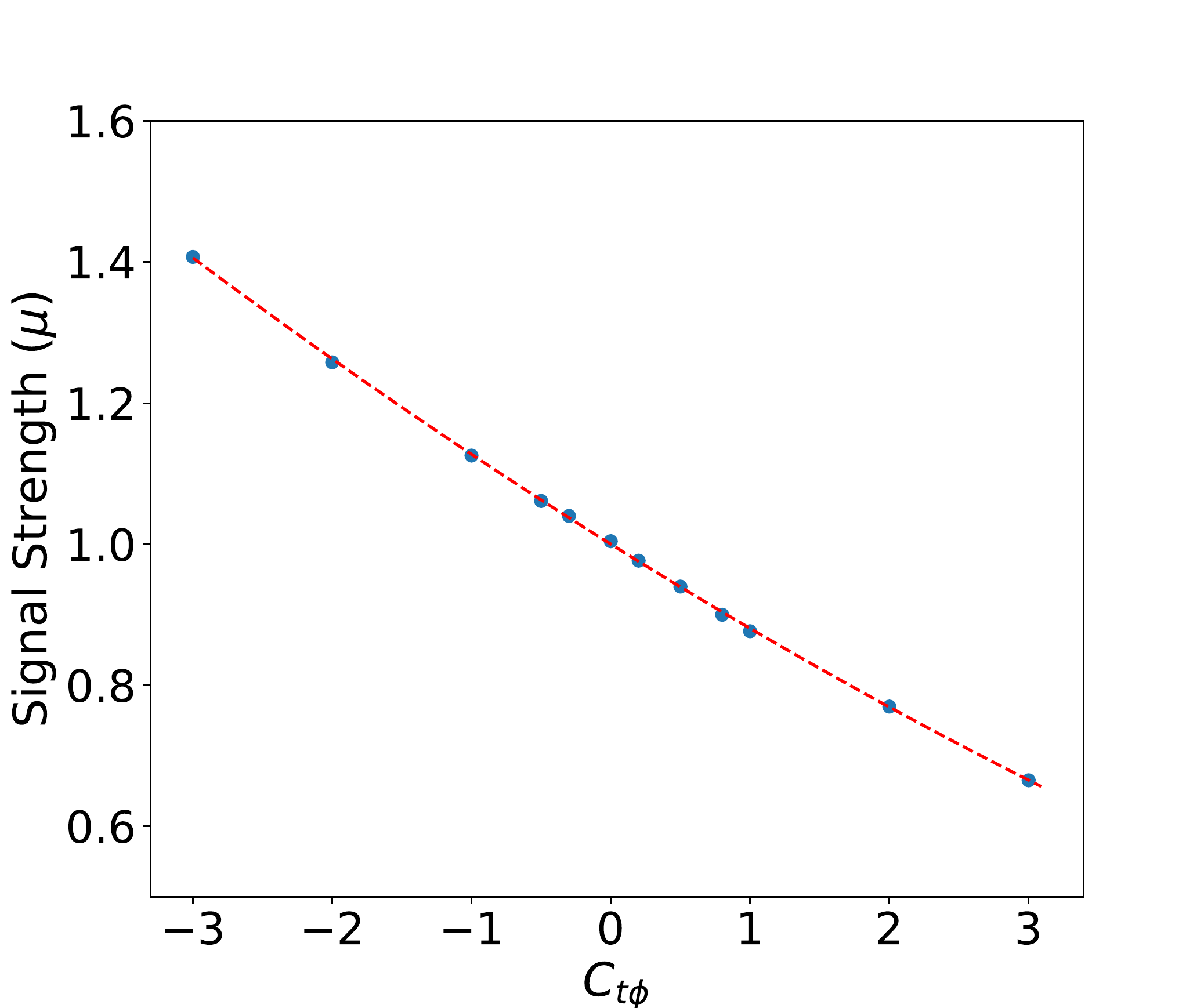}
	\end{subfigure}
	\begin{subfigure}[b]{0.5\textwidth}
		\centering
		\includegraphics[width=8.0 cm]{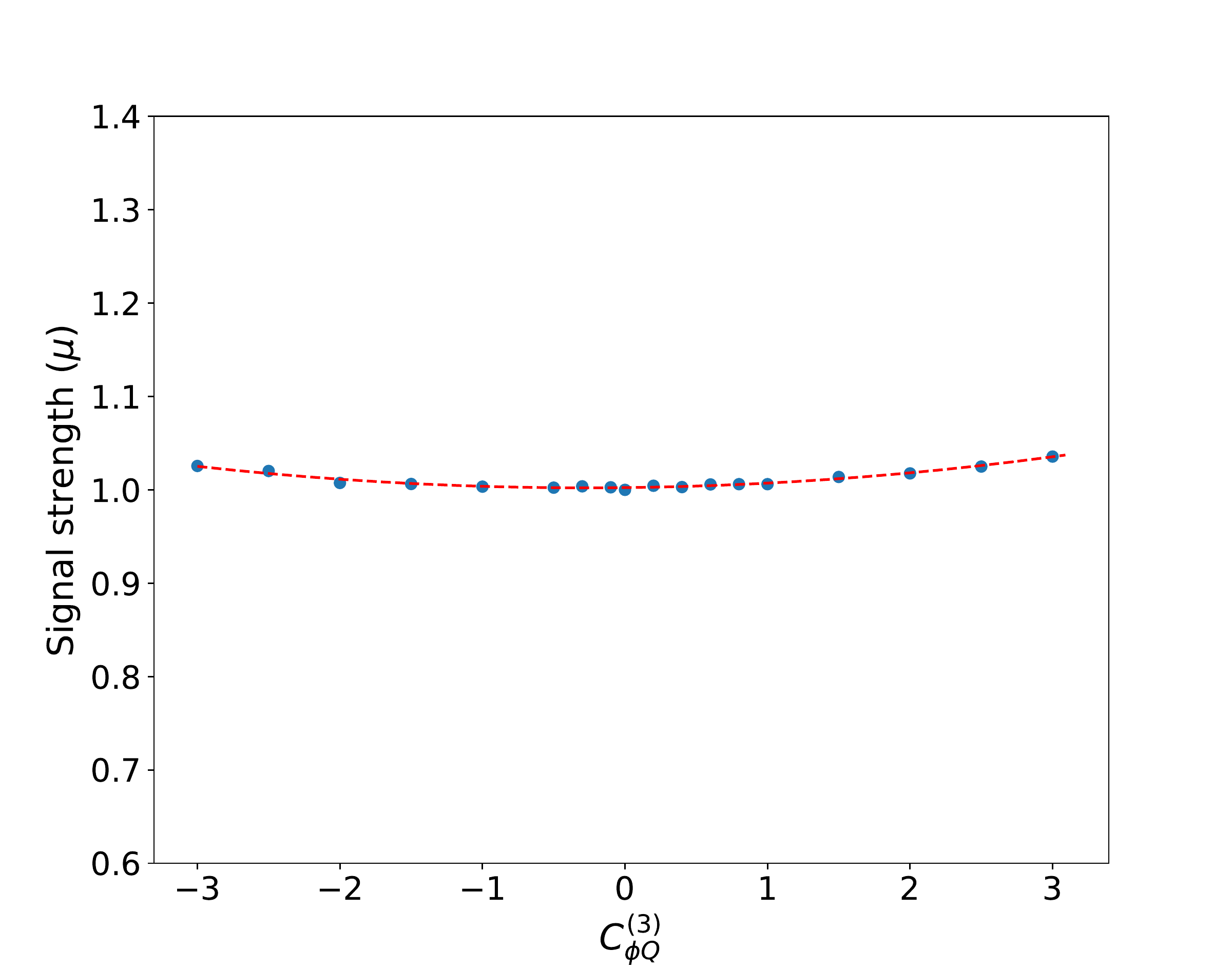}
	\end{subfigure}
	\caption{\small {The signal strength for $\rm t\bar{t}H$
		production for different values of $\rm C_{t\phi}$ (left) and $\rm C_{\phi Q}^{(3)}$ (right), fitted (dashed line) with a quadratic polynomial.}}
	\label{fig:fit3}
\end{figure}

For another illustration, we repeat this exercise by studying the variation of $\rm t\bar{t}H$ signal strength ($\mu$) with the operator $\Ctphi$ or $\Cpq3$ setting all the other WCs to zero.
Fig.~\ref{fig:fit3} presents the variation of $\mu$ with $\Ctphi$(left) and $\Cpq3$(right). The negative slope of the $\mu$ with $\rm \Ctphi$ indicates that the interference term, linear in WC, is the dominant one compared to the quadratic term. The almost constant behavior of the cross-section with $\Cpq3$ implies that the $\rm t\bar{t}H$ process is not much sensitive to this operator.
In these cases, a linear approximation is good enough to fit the $\mu$.

In case of two non-zero SMEFT operators, i.e $n=2$, 
the same procedure is followed by setting two operators of interest to non-zero values, keeping all others to zero. Accordingly, the production cross-sections for a given process can be expressed in terms of two WCs following Eq.~\ref{eq:sigma},
\br
\rm \sigma^{EFT}_{xx}(C_1,C_2)=\sigma^{SM}+\beta_1C_1+\beta_2 C_2+\gamma_{11}C_1^2+\gamma_{22}C_2^2
+2\gamma_{12}C_1 C_2.
\label{eq:sigma_poly}
\er
Note that, the quadratic terms are suppressed by $\Lambda^4$, and hence are not considered in several analysis~\cite{Ellis:2018gqa,daSilvaAlmeida:2018iqo,Ellis:2020unq}.
Following the same strategy as before, the numerical values of these coefficients can be obtained by fitting their 
variation of cross-sections with the corresponding combination of WCs. The procedure in same for other observables, such as, signal strength ($\mu$), asymmetry, etc.
Here, the variation of the $\rm t\bar{t}H$ signal strength with WCs $\Cpq3$ and $\Ctphi$ is fitted with a two dimensional quadratic polynomial. 
The expression of the fitted polynomial the $\rm t\bar{t}H$ signal strength is found to be of the form, 
\br
\mu^{EFT}(\Cpq3,\Ctphi) &=& 1.0 + 0.002~\Cpq3 +0.002~{\Cpq3}^2 -
0.122~\Ctphi \\ \nonumber
&+& 0.004~{\Ctphi}^2 - 0.001~\Cpq3\Ctphi,
\label{eq:c1c2}
\er
where the leading term corresponds to the SM signal strength. The above expression 
indicates that the operator corresponding to the WC $\Ctphi$ has comparatively
more dominant effect on the $\rm t \bar t H$ process. An important point to note here is that,
for any given physical process and center of mass energy, the numerical values of the coefficients of a given term
are unique and independent, i.e, not correlated with the values of other WCs. Repeating
the similar exercise, the numerical coefficients for all the 5 operators (Eq.~\ref{eq:operators}) for the $\rm t\bar t H$ signal strength are obtained.
For the sake of illustration, the numerical values of the coefficients for the linear ($C_i$), quadratic ($C_i^2$) and interference terms ($C_i C_j$) for the tH production cross-section and the $\rm t\bar{t}H$ signal strength  are presented in Table~\ref{tab:coeff_ttz}(a) and Table~\ref{tab:coeff_ttz}(b), respectively. Here the (1,1) entries present the expected SM cross-section for the tH production, whereas other values in the first row represent the values of coefficients for the linear terms. The diagonal entries present the numerical values corresponding to the quadratic terms, while others show the same for the interference terms. We present the Monte-Carlo uncertainties for each of the entries in these tables. For several entries, the uncertainties are much smaller than the values of the coefficients ($\beta_i, \gamma_{jk}$, or $a_i, b_{jk}$, see Eq.~\ref{eq:sigma},~\ref{eq:mu}). The operators corresponding to these entries have a non-zero effect on the considered process. On the other hand, for a few entries, e.g., the interference terms between $C_{Qq}^{(3,8)}$ and other operators in Table~\ref{tab:coeff_ttz}(a), the uncertainties are found to be larger than the values of the coefficients ($\gamma_{jk}$). These can be interpreted as the irrelevance of the corresponding terms to the process, and the non-zero values of the entries arise solely due to uncertainties in the calculation. These tables explicitly show how an individual operator, or two operators in combination, affect a given process. For instance, in the case of $\Ctphi$ in Table~\ref{tab:coeff_ttz}(b), the value of the coefficient for the linear term with a negative sign and smaller value for the same for quadratic term justify the pattern of variation of signal strength shown in Fig.~\ref{fig:fit3} (left). Similarly, the tiny values of coefficients of $\Cpq3$ justify the curve in the same figure (right). Table~\ref{tab:coeff_ttz}(b) also suggests that $C_{Qq}^{(3,1)},\Cqq38$and $\Ctphi$ are comparatively more sensitive to the $\rm t\bar{t}H$ process. On the other hand, Table~\ref{tab:coeff_ttz}(a) indicates that $\Ctw,C_{Qq}^{(3,1)},$ and $\Cpq3$ have the dominant sensitivity to the tH production process.
Following this exercise, eventually, we obtain the WC-dependent functional forms of the cross-section and other observables such as signal strength, differential cross-section, asymmetry etc. for each of the processes considered in this study, precisely what we need in the next step to interpret precision measurements.

\begin{table}
		\caption{\small {The coefficients of various linear ($\beta_i,a_i$) and quadratic terms ($\gamma_{jk},b_{jk}$), following Eq~\ref{eq:sigma} and~\ref{eq:mu}, corresponding to the fits for the cross-section of $\rm tH$ and the signal strength ($\mu$) of $\rm t\bar{t}H$.}}
		\centering  
		\renewcommand{\arraystretch}{1.5}  
		\begin{adjustbox}{width=1.0\textwidth,center}
			\begin{tabular}{c| c c c c c c c}
				\multicolumn{7}{@{}l}{\Large \em(a) Coefficients ($\beta_i,\gamma_{jk}$) for the fits of tH cross-section (units in pb)}\\
				\hline
				\hline
				&1 & $\mathcal{C}_{ t\phi}$  &  $ C_{\phi Q}^{(3)}$ & $ C_{tW}$ & $ C_{Qq}^{(3,1)}$ & $ C_{Qq}^{(3,8)}$ & $C_{\phi W}$\\
				\hline
				1                           & 7.7E-2 $\pm$ 5.0E-5 & $-$3.4E-3 $\pm$ 2.4E-5 & 3.9E-3 $\pm$ 4.0E-5 & 2.2E-2 $\pm$ 8.0E-5 & $-$9.7E-3 $\pm$ 1.4E-4 & 1.9E-4 $\pm$ 8.0E-5& 1.1E-3 $\pm$ 2.6E-5\\
				
				$\mathcal{C}_{ t\phi}$      &  & 9.1E-3 $\pm$ 1.2E-5 & $-$4.0E-3 $\pm$ 2.5E-5 & $-$1.9E-3 $\pm$ 5E-5 & 4.6E-3 $\pm$ 7.0E-5 & $-$1.0E-5$\pm$ 4.0E-5& $-$8.8E-4 $\pm$ 1.5E-5\\	
				
				$ C_{\phi Q}^{(3)}$      &  &  & 1.5E-2 $\pm$ 2.0E-5 & 5.0E-3 $\pm$ 7.0E-5 & $-$1.4E-2 $\pm$ 1.0E-4 &2.0E-5 $\pm$ 5.0E-5& 4.0E-3 $\pm$ 1.7E-5\\
				
				$ C_{tW}$&  &  &  &  6.5E-2 $\pm$ 5.0E-5 & 5.9E-3 $\pm$ 1.6E-4 & 4.0E-4 $ \pm$ 9.0E-5 &1.1E-2 $\pm$ 1.7E-5\\	
				
				$ C_{Qq}^{(3,1)}$    &  &  &  &  & 1.6E-1 $\pm$ 1.1E-4 & 1.0E-5 $\pm$ 1.3E-4& $-$5.0E-3 $\pm$ 7.0E-5\\	
				
				$ C_{Qq}^{(3,8)}$          &  &  &  &  &  & 3.5E-2 $\pm$ 8.0E-5 & 1.2E-5 $\pm$ 3.0E-5\\
				
				$ C_{\phi W}$          &  &  &  &  &  & &6.3E-3 $\pm$ 3.2E-5\\	
				\hline
				\hline
			\end{tabular}
		\end{adjustbox}
	
	\bigskip
	
	\centering    
	\begin{adjustbox}{width=1.0\textwidth,center}
		\begin{tabular}{c| c c c c c c c}
			\multicolumn{7}{@{}l}{\Large \em(b) Coefficients ($a_i,b_{jk}$) for the fits of $t\bar{t}H$ signal strength ($\mu$)}\\
			\hline
			\hline
			&1 & $\mathcal{C}_{ t\phi}$  &  $ C_{\phi Q}^{(3)}$ & $ C_{tW}$ & $ C_{Qq}^{(3,1)}$ & $ C_{Qq}^{(3,8)}$ & $C_{\phi W}$\\
			\hline
			1                           & 1.00$\pm$ 1.1E-4 & $-$1.23E-1 $\pm$ 1.7E-4 & 2.0E-3 $\pm$ 1.7E-4 & 9.2E-3 $\pm$ 1.9E-4 & 1.9E-2 $\pm$ 2.9E-4 & 1.6E-2 $\pm$ 2.1E-4 & 1.1E-3 $\pm$ 1.9E-4\\
			
			$\mathcal{C}_{ t\phi}$      &  & 3.8E-3 $\pm$ 8.0E-5 & $-$8.8E-4 $\pm$ 1.0E-4 & $-$3.4E-4 $\pm$ 1.1E-4 & $-$1.1E-3 $\pm$ 1.5E-4 & $-$7.6E-4 $\pm$ 1.1E-4& $-$1.1E-4 $\pm$ 9.0E-5\\	
			
			$ C_{\phi Q}^{(3)}$      &  &  & 2.6E-3 $\pm$ 7E-5 & $-$8.6E-3 $\pm$ 1.2E-4 & 2.8E-4 $\pm$ 1.7E-4 &1.0E-5 $\pm$ 1.2E-4 & 1.4E-4 $\pm$ 1.1E-4\\
			
			$ C_{tW}$&  &  &  &  2.6E-2 $\pm$ 9.0E-5 & 5.3E-2 $\pm$ 1.9E-4 & 3.0E-5 $\pm$ 1.4E-4 &2.1E-3 $\pm$ 1.2E-4\\	
			
			$ C_{Qq}^{(3,1)}$    &  &  &  &  & 1.26E-1 $\pm$ 1.4E-4 &9.0E-5 $\pm$ 1.9E-4 &7.2E-3 $\pm$ 1.7E-4\\	
			
			$ C_{Qq}^{(3,8)}$          &  &  &  &  &  & 2.8E-2 $\pm$ 1.1E-4 & 6.0E-5 $\pm$ 1.2E-4\\
			
			$ C_{\phi W}$          &  &  &  &  &  & &1.5E-4 $\pm$ 9.0E-5\\	
			\hline
			\hline
		\end{tabular}
	\end{adjustbox}
	\label{tab:coeff_ttz}
\end{table}

\subsection{Experimental measurements}

In this subsection, we present the list of measurements used as inputs in our analysis, to constrain the SMEFT operators.

$\bullet$ {\bf Production of Higgs boson:}
Several Higgs-boson production modes are relevant to the considered set of operators in Eq.~\ref{eq:operators}. For example, the production of Higgs boson in association with vector bosons (WH and ZH) and the vector boson fusion (VBF) include the V-V-H vertex in the production diagram. They are thus crucial to the  SMEFT operator $\Opw$. On the other hand, top-quark production, either in pair or in single, in association with the SM Higgs boson, namely $\rm t\bar{t} H$, tHq, and tHW processes, as shown in Table~\ref{tab:higgsdata}, are sensitive to the rest of the operators in Eq.~\ref{eq:operators}. The measurements are presented in terms of 
signal strength ($\mu$) as defined in Eq~\ref{eq:mu}, including uncertainties and for various luminosity options. 
In case of asymmetric error, the larger value is taken into
consideration. 

\begin{table}[t]
	\caption{\small {Measurements of associated production of single-top and top-pair with Higgs boson, with $\mu$ being the signal strength.}}
	\centering
	\begin{tabular}{c| c c c c c}
		\hline
		\hline
		&       Process & Observable  & $\mathcal{L}~(\invfb)$  &  Measured value & Ref.\\
		\hline
		& $\rm VBF, H\to WW$& $\rm \mu$ & $\;139$ & $0.93^{+0.23}_{-0.20}$ & \cite{ATLAS:2022ooq}\\
		& $\rm VBF, H\to \tau\tau$& $\rm \mu$ & $\;139$ & $0.90^{+0.20}_{-0.17}$ & \cite{ATLAS:2022yrq}\\
		& $\rm VBF, H\to b\bar{b}$& $\rm \mu$ & $\;126$ & $0.95^{+0.33}_{-0.32}$ & \cite{ATLAS:2020bhl}\\
		
		& $\rm ZH, H\to WW$& $\rm \mu$ & $\;139$ & $1.64^{+0.55}_{-0.47}$ & \cite{ATLAS:2022ndd}\\
		ATLAS& $\rm WH, H\to WW$& $\rm \mu$ & $\;139$ & $0.45^{+0.32}_{-0.29}$ & \cite{ATLAS:2022ndd}\\
		& $\rm t\bar{t} H, H\to b\bar{b}$& $\rm \mu$ & $\;79.8$ & $0.79^{+0.61}_{-0.60}$ & \cite{ATLAS:2018mme}\\
		& $\rm t\bar{t} H, H\to ZZ(4\ell)$& $\rm \mu$ & $\;79.8$ &  $\rm <1.77 ~at ~68\% ~CL$ & \cite{ATLAS:2018mme}\\
		& $\rm t\bar{t} H, H\to Multi$-$\rm lepton$& $\rm \mu$ & $\;79.8$ & $1.56^{+0.42}_{-0.40}$ & \cite{ATLAS:2018mme}\\
		\hline
		& $\rm VBF, H\to WW$& $\rm \mu$ & $\;138$ & $0.71^{+0.28}_{-0.25}$ & \cite{CMS:2022uhn}\\
		& $\rm VBF, H\to \gamma\gamma$& $\rm \mu$ & $\;137$ & $1.04^{+0.34}_{-0.31}$ & \cite{CMS:2021kom}\\
		& $\rm ZH, H\to WW$& $\rm \mu$ & $\;138$ & $2.0^{+0.7}_{-0.7}$ & \cite{CMS:2022uhn}\\
		& $\rm ZH, H\to \tau\tau$& $\rm \mu$ & $\;138$ & $2.0^{+0.91}_{-0.81}$(bin 1) & \cite{CMS:2022kdi}\\
		&& &  & $2.18^{+1.0}_{-0.82}$(bin 2) & \\
		
		CMS& $\rm WH, H\to WW$& $\rm \mu$ & $\;138$ & $2.0^{+0.6}_{-0.6}$ & \cite{CMS:2022uhn}\\
		& $\rm WH, H\to \tau\tau$& $\rm \mu$ & $\;138$ & $0.79^{+0.94}_{-0.91}$(bin 1) & \cite{CMS:2022kdi}\\
		&& &  & $2.65^{+1.26}_{-1.15}$(bin 2) & \\
		& $\rm t\bar{t} H, H\to b\bar{b}$& $\rm \mu$ & $\;138$ & $-0.27^{+0.86}_{-0.83}$ & \cite{CMS:2022hjj}\\
		& $\rm t\bar{t} H, H\to b\bar{b}$& $\rm \mu$ & $\;35.9$ & $0.72^{+0.45}_{-0.45}$ & \cite{CMS:2018hnq}\\
		& $\rm t\bar{t} H, H\to ZZ(4\ell)$& $\rm \mu$ &$\;137$ & $\rm 0.16^{+0.98}_{-0.16}$ & \cite{CMS:2021ugl}\\
		& $\rm t\bar{t} H, H\to Multi$-$\rm lepton$& $\rm \mu$ & $\;35.9$ & $\rm 1.23^{+0.45}_{-0.43}$ & \cite{CMS:2018fdh}\\
		& $\rm tHq+tHW, combined$& $\rm \sigma$ & $\;35.9$ & $\rm 0.92^{+0.40}_{-0.27}$ pb & \cite{CMS:2018jeh}\\
		\hline
		\hline
	\end{tabular}
	\label{tab:higgsdata}
\end{table}

$\bullet$ {\bf Single top-quark production:}
This set of data includes only the single top-quark production along
with a jet in the t-channel process. However, the single top quark production in
association with a W boson
(tW production) is treated in a different category.
These measurements are mostly sensitive to $C_{\phi Q}^{(3)}$
and $C_{tW}$ via the W-t-b vertex, as seen in Table~\ref{tab:fishinfo}.
The measurements include mainly the total production cross-sections and as well as normalized
differential cross-sections, listed in Table~\ref{tab:singletop}. The added advantage of using
normalized differential cross-sections is to have the minimal effect of uncertainties in comparing the theoretical predictions (LO) to the observed one. Although there are several differential cross-section measurements corresponding to different observables, in this study, we have considered only one ($\frac{1}{\sigma}\frac{ d\sigma}{d|y|}$) of them, since others must be highly correlated.

\begin{table}[t]
	\caption{\small {Measurements of single top-quark production and normalized differential cross-sections, in terms of the rapidity (y) of the reconstructed top-quark.}}
	\centering
	\begin{adjustbox}{max width=\textwidth}
		\begin{tabular}{c| c c c c c}
			\hline
			\hline
			&       Process & Observable  &  $\rm\mathcal{L}~(\invfb)$ & Measured value & Ref.\\
			\hline
			& tj (t-channel)& $\rm \sigma_{tot}(t)$ & $\;35.9$ & 130$\pm 19$ pb & \cite{CMS:2018lgn}\\
			&tj (t-channel)& $\rm \sigma_{tot}(\rm \bar{t})$ & $\;35.9$ & 77$\pm 12$ pb & \cite{CMS:2018lgn}\\
			CMS & tj (t-channel)& $(1/\sigma) d\sigma/d|y^{t+\bar{t}}|$ & $\;2.3$&  $0.64\pm 0.14$(bin 1) & \cite{CMS:2016xnv}\\
			& &  & & $0.55\pm 0.12$(bin 2) & \\
			& &  & & $0.50\pm 0.12$(bin 3) & \\
			& &  & & $0.18\pm 0.08$(bin 4) & \\
			&& $(1/\sigma) d\sigma/d|y^{t}|$ & $\;35.9$ & $0.58\pm 0.15$(bin 5)  & \cite{CMS:2019jjp}\\
			& & & &$0.53\pm 0.08$(bin 6)\\
			
			& & & &$0.50\pm 0.09$(bin 7)\\
			& & & &$0.47\pm 0.09$(bin 8)\\
			& & & &$0.26\pm 0.02$(bin 9)\\
			\hline
			ATLAS& tj (t-channel)& $\rm \sigma_{tot}(t)$ & $\;3.2$ & 156$\pm 28$ pb & \cite{ATLAS:2016qhd}\\
			&tj (t-channel)& $\rm \sigma_{tot}(\rm \bar{t})$ & $\;3.2$ & 91$\pm 19$ pb & \cite{ATLAS:2016qhd} \\
			\hline
			\hline
		\end{tabular}
	\end{adjustbox}
	\label{tab:singletop}
\end{table}

$\bullet$ {\bf Top-quark pair production and $\rm t\bar{t}$ asymmetry:}
There are a good number of measurements of the top-quark pair production and the $\rm t\bar{t}$ asymmetry, but here we focus only on some most sensitive recent measurements (see Table \ref{tab:tt}). We particularly used the differential cross-section measurements for $\rm t\bar{t}$. Though differential distributions in various kinematic bins are measured, we considered one of them from each analysis to avoid unknown experimental correlations between pairs of distributions, which might lead to double counting.

These measurements are mostly sensitive to the four-fermi operators ($C_{Qq}^{(3,1)}$ and $\Cqq38$) (see Table~\ref{tab:fishinfo}).

\begin{table}[t]
	\caption{\small {Measurements of $\rm t\bar{t}$ charge-asymmetry ($A_C$) and normalized top-quark pair production differential cross-sections, in terms of the invariant mass ($\rm m_{t\bar{t}}$) of the top pair. }}
	\centering
	\begin{adjustbox}{max width=\textwidth}
		\begin{tabular}{c| c c c c c}
			\hline
			\hline
			&       Process & Observable  &  $\rm\mathcal{L}~(\invfb)$ & Measured value & Ref.\\
			\hline
			ATLAS & $\rm t\bar{t}$ & $(1/\sigma) d\sigma/dm_{t\bar{t}}$ & $\;139$&  $0.21\pm 0.0023$(bin 3) & \cite{ATLAS:2022xfj}\\
			& &  & & $0.25\pm 0.003$(bin 4) & \\
			& &  & & $0.18\pm 0.0035$(bin 5) & \\
			& &  & & $0.11\pm 0.0044$(bin 6) & \\
			\hline   
			CMS & & $(1/\sigma) d\sigma/dm_{t\bar{t}}$ & $\;137$ & $0.10\pm 0.15$(bin 1)  & \cite{CMS:2022uae}\\
			& & & &$0.16\pm 0.08$(bin 2)&\\
			
			& & & &$0.21\pm 0.09$(bin 3)&\\
			& & & &$0.20\pm 0.09$(bin 4)&\\
			& & & &$0.14\pm 0.02$(bin 5)&\\
			& & & &$0.08\pm 0.02$(bin 6)&\\
			& & & &$0.05\pm 0.02$(bin 7)&\\
			& & & &$0.28\pm 0.02$(bin 8)&\\
			\hline
			ATLAS& $\rm t\bar{t}$ & $A_C$ & $\;139$ & 0.0060$\pm 0.0015$ & \cite{ATLAS-CONF-2019-026}\\
			\hline
			CMS&$\rm t\bar{t}$& $A_C$ & $\;138$ & 0.69$^{+0.65}_{-0.69}$ & \cite{CMS:2022ged} \\
			\hline
			\hline
		\end{tabular}
	\end{adjustbox}
	\label{tab:tt}
\end{table}

$\bullet$ {\bf Top-quark pair production in association with a vector boson:}
In this category of measurements, $\rm t\bar{t}Z$ and $\rm t\bar{t}W$ processes
are considered. It includes the measurements of total inclusive cross-sections
and differential cross-section like single top case, as shown in Table~\ref{tab:ttv}. This set of data is primarily
useful in constraining $C_{\phi Q}^{(-)}$ and $C_{\phi t}$ (see table \ref{tab:fishinfo}).
For both the processes, the combined measurements with the final states
consisting of $3\ell$  and $4\ell$ are considered and the corresponding theoretical cross
sections are also estimated according to those final states.

\begin{table}[t]
	\caption{\small {Measurements of associated production of top-quark pair with a vector boson.}}
	\centering
	
	\begin{tabular}{c| c c c c c}
		\hline
		\hline 
		&       Process & Observable  &  $\rm \mathcal{L}~(\invfb)$ & Measured value & Ref.\\
		\hline
		& $\rm t\bar{t}Z$& $\rm \sigma_{tot}$ &$\;36.1$ & 0.95$\pm 0.13$ pb & \cite{ATLAS:2019fwo} \\
		&$\rm t\bar{t}W$& $\rm \sigma_{tot}$ &$\;36.1$ & 0.87$\pm 0.19$ pb & \cite{ATLAS:2019fwo} \\
		& $\rm t\bar{t}Z$& $(1/\sigma) d\sigma/dp_T^Z$ &$\;139$ & $0.0018\pm 0.0013$(bin 1) & \cite{ATLAS:2021fzm}\\
		& & & &$0.0055\pm 0.0025$(bin 2)  & \\
		ATLAS & & & &$0.0053\pm 0.002$(bin 3) & \\
		& & & &$0.0057\pm 0.0015$(bin 4)  & \\
		& & & &$0.0022\pm 0.00085$(bin 5) & \\
		& & & &$0.0006\pm 0.0004$(bin 6)   & \\
		& & & &$0.0006\pm 0.00025$(bin 7))  & \\
		\hline
		& $\rm t\bar{t}Z$& $\rm \mu$ & $\;138$ & $0.65^{+1.04}_{-0.98}$ & \cite{CMS:2022hjj}\\
		& $\rm t\bar{t}Z$& $\rm \sigma_{tot}$ &$\;35.9$ & 0.99$\pm 0.14$ pb & \cite{CMS:2017ugv}\\
		&$\rm t\bar{t}W$& $\rm \sigma_{tot}$ &$\;35.9$ & 0.77$\pm 0.17$ pb & \cite{CMS:2017ugv} \\
		& $\rm t\bar{t}Z$& $\rm \sigma_{tot}$ &$\;77.5$ & 0.95$\pm 0.08$ pb & \cite{CMS:2019too} \\
		CMS & $\rm t\bar{t}Z$& $(1/\sigma) d\sigma/dp_T^Z$ &$\;77.5$ & $0.004\pm 0.001$(bin 1) & \cite{CMS:2019too}\\
		& & & &$0.005\pm 0.0009$(bin 2) &  \\
		& & & &$0.0022\pm 0.0005$(bin 3)& \\
		& & & &$0.0003\pm 0.0001$(bin 4) & \\
		\hline
		\hline
	\end{tabular}
	\label{tab:ttv}
\end{table}
$\bullet$ {\bf Single top-quark production in association 
	with vector boson:}
This category includes $\rm tZq$ and tW production both in ATLAS and CMS experiments and the measurements of the total and differential cross-sections are taken into account. The lists of measurements used in the analysis are shown in Table \ref{tab:tv}. According to Table~\ref{tab:fishinfo}, these measurements mainly restrict $C_{\phi Q}^{(3)}$.

\begin{table}[t]
	\caption{Measurements of associated production of single top-quark with vector boson.}
	\centering
	\begin{tabular}{c| c c c c c}
		\hline
		\hline
		&	Process & Observable  &  $\rm \mathcal{L}~(\invfb)$ & Measured value & Ref.\\
		\hline
		& $\rm tZq(l^+,2\ell)$& $\rm \sigma_{tot}$ & $\;138$& $62.2^{+7.4}_{-6.8}$ fb & \cite{CMS:2021ugv}\\
		CMS	& $\rm tZq(l^-,2\ell)$& $\rm \sigma_{tot}$ &$\;138$ & $26.1^{+5.6}_{-5.4}$ fb & \cite{CMS:2021ugv}\\
		&$\rm tW$& $\rm \sigma_{tot}$ &$\;36$ & 89$\pm 13$ pb & \cite{CMS:2021vqm}\\
		& $\rm tW$& $\rm \sigma_{tot}$ & $\;35.9$& 63$\pm 7$ pb & \cite{CMS:2018amb}\\
		& $\rm tZ$& $ 1/\sigma d\sigma/dp_T^Z$&$\;138$ & $0.007\pm 0.002$(bin 1) & \cite{CMS:2021ugv}\\
		& & & &$0.002\pm 0.007$(bin 2)   & \\
		& & & &$0.0026\pm 0.001$(bin 3)& \\
		& & & &$0.0008\pm 0.0004$(bin 4) & \\
		\hline
		ATLAS& $\rm tZ$& $\rm \sigma_{tot}$ &$\;139$ &  97$\pm 15$ pb & \cite{ATLAS:2020bhu}\\
		&$\rm tW$& $\rm \sigma_{tot}$ & $\;3.2$& $94^{+30}_{-24}$ pb & \cite{ATLAS:2016ofl}\\
		\hline
		\hline
	\end{tabular}
	\label{tab:tv}
\end{table}

\subsection{Fisher information matrix}
In order to have a quantitative assessment of the level sensitivity
of these SMEFT operators to various processes,
we resort to the formalism of the FIM~\cite{Brehmer:2016nyr}.
The FIM provides a quantitative comparison
of sensitivity of a given set of SMEFT operators among a set of processes.

Defining $f(X|\vec{C})$ as the distribution of the experimental
measurements X and the given
true values of the set of WCs $\vec C$,  the FIM
is defined as \cite{Brehmer:2016nyr}
\br
I_{ij}(\vec{C})=-E\left[\frac{\partial^2 log f(X|\vec{C})}
{\partial C_{i}\partial C_{j}}\right],
\label{eq:fim}
\er
where `E' represents the expectation value over a set of
measurements X,  and indices $i,j$ stand for the labels of WCs.
One can argue that
the smallest uncertainty for the WC ($C_i$) can be obtained using the Cramer-Rao bound on the covariance matrix $\mathrm{Cov}_{ij}(\vec{C})$~\cite{Rao1992},
\br
\mathrm{Cov}_{ij}\geq (I^{-1})_{ij}.
\er
If there are $\rm N_{exp}$ number of measurements of an
observable `X', which depend
on the $\rm N_{WC}$ number of WCs, then assuming
$f(X|\vec{C})$ to be Gaussian like as,
\br
f(X|\vec{C}) = \prod_{m=1}^{N_{exp}}\frac{1}{\sqrt{2\pi \delta_{exp,m}^2}}exp\left[-
\frac{\left(X_m^{\mathrm{exp}}-X_m^{\mathrm{EFT}}(\vec{C})\right)^2}{2\delta_{exp,m}^2}\right],
\er
the expanded form of FIM can be constructed following
Eq.~\ref{eq:fim}~\cite{Ethier:2021bye},
\br
\resizebox{0.96\textwidth}{!}
{
	$I_{ij}=E\left[\sum_{m=1}^{N_{exp}}\frac{1}{\delta^2_{exp,m}}
	\left(\gamma_{m,ij}\left(X_m^{\mathrm{EFT}}-X_m^{\mathrm{exp}}\right)
	+\left(\beta_{m,i}
	+\sum_{l=1}^{N_{WC}}C_l \gamma_{m,il}\right)
	\left(\beta_{m,j}+\sum_{l'=1}
	^{N_{WC}}C_{l'} \gamma_{m,jl'}\right)\right)\right]$.~~~~}
\er
Here $\delta_{exp}$ is the experimental uncertainty, $\beta$ and $\gamma$
are the numerical values of the linear and quadratic coefficients of the theoretical expression of observable X respectively (see, e.g., Eq.~\ref{eq:sigma} when $\rm X^{EFT}\equiv\sigma^{EFT}$). Now, 
following the procedure described in Ref.~\cite{Brehmer:2016nyr},
we define a six-dimensional basis of WCs as,
\br 
\vec{C}\equiv\left(\Cpq3~~\Cqq38~~C_{Qq}^{(3,1)}~~\Ctphi~~\Ctw ~~\Cpw\right)^T.
\label{eq:wc_vector}
\er
With this choice, $ I_{ij}$ is evaluated for a set of processes,
such as $\rm tH, t\bar{t}H$, single-top, $\rm t\bar{t}V (V=W,Z),tZq$ and tW.
After diagonalization of $I_{ij}$, the  diagonal entries
quantify the level of sensitivity of
the $i^{th}$ WC ($ C_i$) to a given physical process.
In other words, those present the values of sensitivity of a given process to
the $i^{th}$ WC. This quantification indicates the relative strength by which the $i^{th}$
WC can be constrained from the measurements of that process.
For the sake of illustration, the FIM for the set of $\vec{C}$~(Eq.~\ref{eq:wc_vector}) considering
the single-top (tj) production and differential cross-section measurements (see Table~\ref{tab:singletop}) is given by,
\begin{center}
	\(
	\begin{bmatrix}
	9.52& 0.27 & ~-11.20 &~0 & 3.25 &~0 \\
	0.27 & 89.29 & ~2.63 & ~0 & 0.48 &~0\\
	-11.20 & 2.64   & ~588.81 &~ 0 & 12.38&~0\\
	0 & 0 & ~0 & ~0 & 0 &~0\\
	3.25 & 0.48 & ~12.38 & ~0 & 20.20&~0\\
	0 & 0 & ~0 & ~0 & 0&~0\\
	\end{bmatrix}
	\).
\end{center}
The corresponding eigenvalues 
are $\mathrm{diag}(I_{ij})=\{ 8.30, 89.28, 589.27, 0, 20.96,0\}$. The set of diagonal entries of diagonalized $I_{ij}$ matrix
clearly indicates that $C_{Qq}^{(3,1)}$ is most sensitive to single-top
production,
followed by $\Cqq38$.
The effect of other WCs, such as, $\Cpq3$ and $\Ctw$ are almost negligible.
Since, only the relative size of the entries of the $I_{ij}$ is important, not the absolute size, we normalize
these entries in such a way that $\mathrm{\sum_{i=1}^{N_{WC}} |diag}(I_{ii})|=100$. It gives a normalized measure
of the relative sensitivity of the WCs for a given process. Following the above
strategy, the FIM for all our considered processes and operators are derived and presented in Table~\ref{tab:fishinfo}(a). On the other hand, relative sensitivities of different processes to a given operator depend on the relative sizes of the same diagonal entries (say, $(k,k)$, where $k=1,..,6$, with (1,1) for $\Cpq3$, (2,2) for $\Cqq38$, and so on.) for the FIMs corresponding to respective processes. Thus, in this case, we normalize entries for individual operators over all processes, such as $\mathrm{\sum_{n=1}^{N_{proc}} |diag}(I_{kk})|^{(n)}=100$. Following this exercise, we present the sensitivities of an operator to different processes in Table~\ref{tab:fishinfo}(b).

\begin{table}
	\caption{\small {The relative sensitivity of WCs to various processes after normalizing to the scale of 100, and computed using measurements from Table~\ref{tab:higgsdata}-\ref{tab:tv}. Entries below $1\%$ are dropped, except for the entries of the tH process. Also, $\rm V\equiv W,Z$ wherever relevant.} }
	\centering
	\begin{tabular}{c c c| c| c| c| c| c| c| c| c | c | c}
		\multicolumn{12}{@{}l}{\em(a) Normalization performed for individual processes.}\\
		\hline
		\hline
		& WC &tH&$\rm t\bar{t}H$& VBF & ZH & WH &tj &$\rm t\bar{t}V$ &tZ &  tW & AC & $\rm t\bar{t}$\\
		\hline
		\hline
		&$\Cpq3$&14.5 & - & - & -& -& 1.2 & - & 1.1& 52.5& - &-\\
		&$\Cqq38$&2.2 & 13.0 &- & - & -&12.5 &11.8 &11.7& -& 77.1 & 32.3\\
		&$C_{Qq}^{(3,1)}$& 11.1 &81.1 &- & -&- &83.2 &88.1 &81.0& -& 22.5 &67.1\\
		&$\Ctphi$ &1.3 &5.7 & -& -& -&- &- &-& -& - &-\\
		&$\Ctw$& 70.6 &- &- && -- &2.9 &- &5.6& 47.3& - &-\\
		&$C_{\phi W}$ & 0.3 &- & 100&100& 100 &- &- &-& -& - &-\\
		\hline
		\hline
	\end{tabular}

\bigskip
\begin{tabular}{c c c c c c c c c  c c c c}
	\multicolumn{12}{@{}l}{\em(b) Normalization performed for individual operators.}\\
	\hline
	\hline
	& WC &tH&$\rm t\bar{t}H$ & VBF & ZH &  WH &tj &$\rm t\bar{t}V$ &tZ & tW & AC & $\rm t\bar{t}$\\
	\hline
	\hline
	&$\Cpq3$&68.7 &- &- &- &- &4.4 &- & 14.4 &11.9  & -& -\\
	\hline
	&$\Cqq38$&1.6 &- &- &- &- &6.6 &5.2 & 40.5 &- &30.4 &15.1 \\
	\hline
	&$C_{Qq}^{(3,1)}$&1.0 &- &- &- &- &5.1 &4.9 &34.2 &- &54.1 & -\\
	\hline
	&$\Ctphi$ &74.1 &24.8 &- &- &- &- &- &- &- &- &- \\
	\hline
	&$\Ctw$&71.9 &- &- &- &- &1.9 &- &22.9 &3.3 &- &- \\
	\hline
	&$C_{\phi W}$&0.5 &- &31.8 &26.9 &40.7 &- &- &- &- &- &- \\
	\hline
	\hline
\end{tabular}
	\label{tab:fishinfo}
\end{table}
Each column of the Table~\ref{tab:fishinfo}(a) presents the relative sensitivities of the set of WCs corresponding to respective physics processes. For instance, the operator corresponding to WCs $C_{Qq}^{(3,1)}$ and $C_{Qq}^{(3,8)}$ are primarily sensitive to all processes except tW production, where the 4-F vertex is not present. On the other hand, Table~\ref{tab:fishinfo}(b) helps understand the effects of different processes on individual operators. This table shows that the tH measurement is more effective in constraining the operators $\Opqthree, \Otw$, and $\Otphi$ than other processes. However, it does not significantly impact the 4-Fermi and the $\mathcal{O}_{\phi W}$, where other Higgs boson or top-quark-related measurements are more effective.
 The information provided by this table are very useful in selecting the set of dataset to constrain a given set of SMEFT operators.
This exercise leads us to conclude that
the measurements related to the top quark production
either in single or in pair, in association with Higgs and vector bosons
are very useful to constrain the set of operators considered in this analysis.

\subsection{Constraining WCs}
In the earlier sections, we have laid out the theoretical setup and discussed in detail the level of sensitivity of each of the SMEFT operators (Eq.~\ref{eq:fim}) with the processes involving top-quark (see Table \ref{tab:fishinfo}). Accordingly, the list of precision measurements to be used as input is also identified. With all these settings, now we are set to find the constrained values of these six WCs by the method of $\chi^2$-minimization.

For a given set of WCs, $\vec{C}$, the $\chi^2$ function is defined as,
\br
\chi^2 (\vec{C})=\frac{1}{\mathrm{N_{dat}}}\sum_{i,j=1}^{\mathrm{N_{dat}}}
\left({\cal O}^{exp}_{i}-{\cal O}^{th}_{i}(\vec{C})\right)\left[\mathrm{Cov}_{ij}\right]^{-1}
\left({\cal O}^{exp}_{j}-
{\cal O}^{th}_{j}(\vec{C})\right),
\er
where, ${\cal O}^{\rm exp}_{i}$ and ${\cal O}^{\rm th}_{i}$ are 
the measured and theoretical values of the i-th observables, 
which can be total or differential cross-section, signal strength etc. Here $\rm N_{dat}$ represents the total number of data used in the $\rm \chi^2$-minimization.
The covariance matrix ${\rm Cov}^{-1}_{ij}$ contains the experimental and 
theoretical uncertainties. The $\rm \chi^2$-minimization is performed  
using the framework of TMinuit with the option MIGRAD~\cite{James:1994vla}. 

The convergence of MIGRAD method of $\chi^2$-minimization 
was confirmed, and the stability of the best-fit values 
are also checked with respect to different initial conditions 
of the WCs and the step sizes. The quality of $\chi^2$-minimization 
is ensured by verifying the quantity $\rm \chi^2/N_{dat}$, which turned 
out to be 0.9 (1.4) in case of linear (quadratic) fit for a total number of data, 
$\rm N_{dat}=70$.

\begin{figure}[t]
	\centering
	\includegraphics[width=8.5 cm]{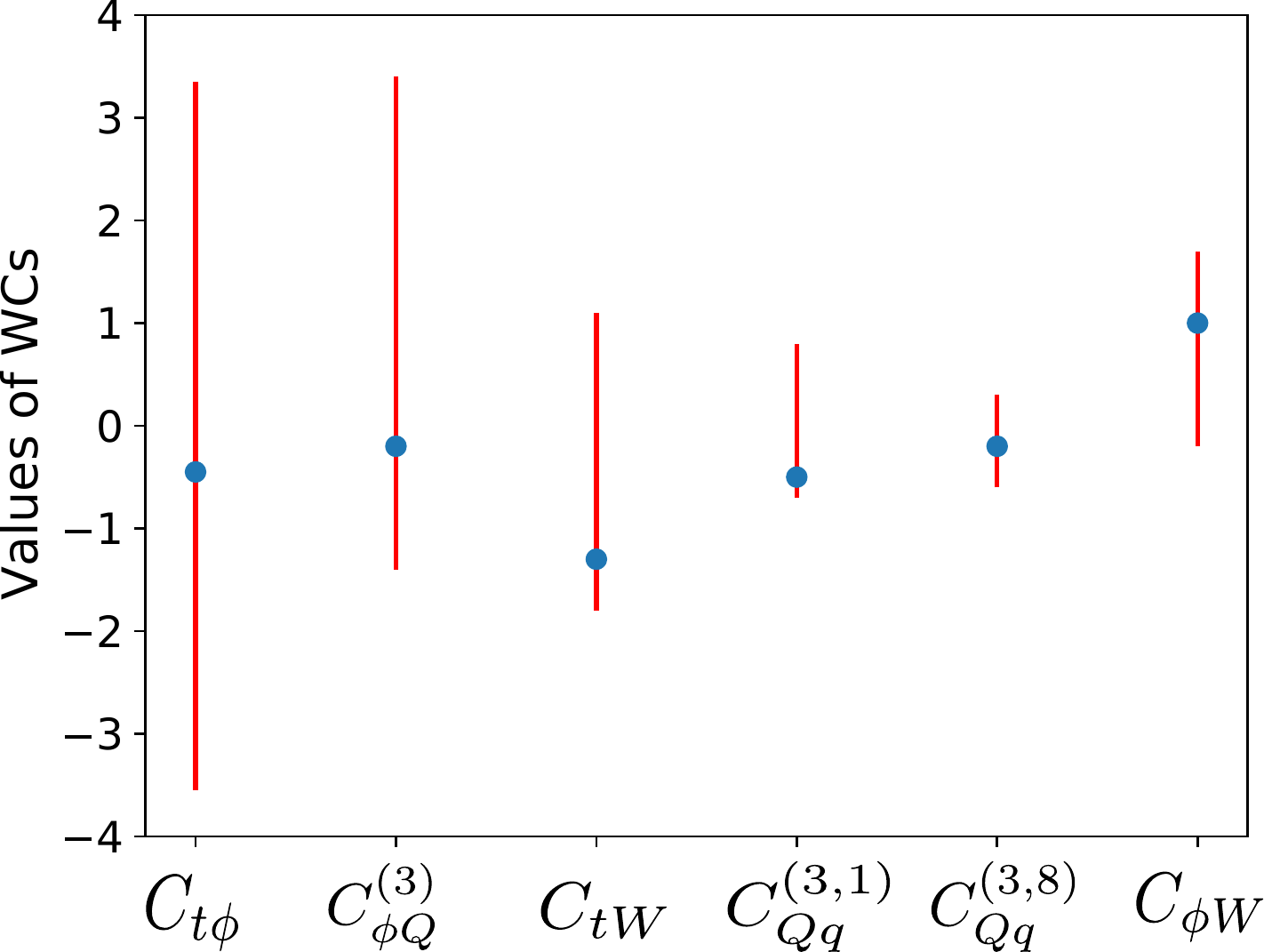}
	\caption{\small {Best-fit values and corresponding allowed ranges for the WC, from combined quadratic fit (Table \ref{tab:quad}).}}
	\label{fig:bf-ranges}
\end{figure}

We perform the fitting for two cases.
In the first case, a given WC is fitted setting others to zero and label it as
``single fit''. Whereas, for the second case, fitting is carried out 
for all the operators, 
calling it a ``combined fit''. Depending on the order of the used theoretical values, the fitting is performed both at linear and quadratic level (as shown in Eq~\ref{eq:sigma}).
The results for both linear and quadratic fits are presented in 
Table \ref{tab:lin2} and Table \ref{tab:quad} respectively, in terms of the best-fit values and the corresponding $2\sigma$ error. In each case, both the single fit and the combined fit scenarios are considered. For the sake of presentation, the best-fit values of the combined quadratic fit are also shown in Fig.~\ref{fig:bf-ranges}, along with $2\sigma$ error bands.

\begin{table}[t]
	\caption{Linear fit of WCs, for both the single and the combined fit, using the full dataset Table~\ref{tab:higgsdata}-\ref{tab:tv}.}
	\centering
	\begin{tabular}{c|cccccc}
			\hline
			\hline
		& $ C_{t\phi}$ & $ C_{\phi Q}^{(3)}$ & $ C_{tW}$ & $ C_{Qq}^{(3,1)}$ & $ C_{Qq}^{(3,8)}$& $ C_{\phi W}$\\
		\hline
		Single fit  & -1.0$^{+1.7}_{-1.7}$ & 2.0$^{+1.3}_{-1.3}$ & 0.8$^{+2.2}_{-2.2}$ & -0.3$^{+0.9}_{-0.9}$ & 0.5$^{+1.5}_{-1.4}$ & 1.3$^{+0.6}_{-0.6}$\\
		\hline
		Combined fit   & -1.9$^{+3.4}_{-3.4}$ & 2.2$^{+2.2}_{-2.3}$ & 2.0$^{+3.0}_{-3.0}$ & -0.4$^{+0.45}_{-0.45}$ & 0.7$^{+1.5}_{-1.4}$&1.2$^{+0.6}_{-0.6}$\\
		\hline
		\hline
	\end{tabular}
	\label{tab:lin2}
\end{table}

\begin{table}[t]
	\caption{Same as Table~\ref{tab:lin2}, but for quadratic fit of WCs.}
	\centering
	\begin{tabular}{c|cccccc}
		\hline
		\hline
		& $ C_{t\phi}$ & $ C_{\phi Q}^{(3)}$ & $ C_{tW}$ & $ C_{Qq}^{(3,1)}$ & $ C_{Qq}^{(3,8)}$ & $ C_{\phi W}$\\
		\hline
		Single fit  & -1.6$^{+3.7}_{-2.9}$ & 2.0$^{+1.0}_{-1.2}$ & 1.2$^{+0.5}_{-1.1}$ & -0.2$^{+0.5}_{-0.5}$ & -0.5$^{+0.4}_{-0.3}$& 1.1$^{+0.7}_{-1.1}$\\
		\hline
		Combined fit  & -0.45$^{+3.8}_{-3.1}$ & -0.2$^{+3.4}_{-1.8}$ & -1.3$^{+2.6}_{-0.7}$ & -0.5$^{+1.3}_{-0.2}$ & -0.2$^{+0.5}_{-0.4}$&1.0$^{+0.7}_{-1.2}$\\
		\hline
		\hline
	\end{tabular}
	\label{tab:quad}
\end{table}
Evidently, in the quadratic fitting, the magnitudes of the central values of the constrained WCs are reduced significantly, while the error bands remain almost unchanged. An exact matching between the theoretical prediction and the corresponding experimental measurements is expected to constrain WC severely, whereas, a wide gap between them may not be effective to put any tight constraint on them. In this regard, Table~\ref{tab:fishinfo} predicts the set of datasets that can potentially constrain a subset of operators.        

The systematics in the measured values translate the error in the fitted values of the WCs. For example, the $\Ctw$ operator affecting the vertex W-t-b is expected to be very sensitive to tH, tZ, and tW production, as seen in Table~\ref{tab:fishinfo}. Therefore, the measurements of cross-section, signal strength, or differential cross-section in various channels of these processes, as shown in Table~\ref{tab:higgsdata}, and \ref{tab:tv}, primarily constrain $\Ctw$ operator. Comparatively larger uncertainties in these measurements result in a substantial error to the best-fit value of $\Ctw$. In contrast, 4-F operators are much more tightly constrained due to the measurements from tj, $\rm t\bar{t}V$, $\rm t\bar{t}$-$\rm A_C$, etc. Following similar arguments and considering Table~\ref{tab:higgsdata}-\ref{tab:tv}, one can have some understanding of the best-fit values of the WCs as presented in Table~\ref{tab:quad}.
The uncertainties of theoretical cross-sections are not taken into consideration while performing fitting. However, we have checked that, a flat $10\%$ theoretical uncertainty of all the observables affects the best-fit values of the WC with a shift by around $7-8\%$.

\section{Implications at the LHC}
The previous section presents the best-fit values within a constrained range of operators relevant to tHq process. The next goal is to explore the feasibility of finding the signal of these operators at the LHC with the present and future luminosity options. As mentioned before, the signal, if exists, is expected to be observed at the tail of the kinematic distribution of certain chosen observables constructed out of the momenta of final state particles of the tHq process. In this case, the primary reason for such possible deviations can be attributed to the energy growth in the scattering sub-amplitude $\rm bW\to tH$~\cite{Farina:2012xp}, where H is radiated off either from t or W boson.

\begin{figure}[H]
	\centering
	\includegraphics[width=5.5 cm]{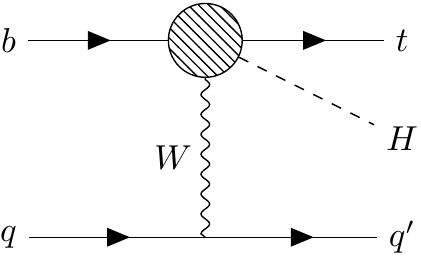}
	\caption{\small {Representative Feynman diagram of tHq process, where the scattering sub-amplitude $\rm bW\to tH$ can be embedded in the blob.}}
	\label{fig:diag-thq}
\end{figure}
The presence of this sub-amplitude $\rm bW\to tH$, in the tHq production process is shown in Fig.~\ref{fig:diag-thq}. Similar effects are also present in tZj, tWZ, and other production processes involving gauge or Higgs bosons, where different other scattering sub-amplitudes
contribute to the energy growth~\cite{Degrande:2018fog,Maltoni:2019aot,Faham:2021vde}. On the contrary, these kinds of sub-amplitudes do not exist in the single-top, $\rm t\bar{t}$, $\rm t\bar{t}H$ or $\rm t\bar{t}V$ processes, implying the absence of any energy growth due to the effect of the SMEFT
operators. Consequently, while considering tHq process as the signal, even if these SMEFT operators may affect some of the dominating background processes, especially the $\rm t\bar{t}H$ production, the deviations in the kinematic distributions of the tHq process are expected to be visible over the background. Therefore, despite having a very low cross-section, the tHq production with a dedicated analysis strategy may turn out to be more promising compared to other common processes in probing the SMEFT effects at the LHC.

As already mentioned, the impact of the SMEFT operators on any process is very likely to be visible at the higher side of the kinematic distribution, supposed to be the signal region. With this understanding, we perform the analysis by splitting the phase space into two regions, 
(a) boosted region, with  $\rm p_T (H)>300~GeV$, and (b) 
non-boosted region, with $\rm p_T (H)<300~GeV$. 
Additionally, both the hadronic and the leptonic decay channels of top-quark are 
considered separately.  
In the final state, the Higgs boson is reconstructed out of two b-jets, whereas the top quark is reconstructed as a top-jet for its hadronic decay mode only, for both the scenarios (a) and (b).
In addition to the reconstructed top-quark and Higgs boson, the presence at least one non-tagged jet in the final state is required for both the cases. Additionally, a tagged b-jet is required for the leptonic final state. Thus, our signal final states are categorized as,

\br
\rm H_{ reco}+ t_{ reco}+ n\mhyphen jets~~~~(hadronic~ final~ state)
\label{eq:signal1}\\
\rm ~~~~~~H_{ reco}+\ell+ n_b\mhyphen jets + n\mhyphen jets ~~~~(semi-leptonic~ final~ state),
\label{eq:signal2}
\er
with $n_b,n \geq 1$. 	
The dominant SM background contributions are expected to be due to the following processes,
\br
\rm p~p \rightarrow t\bar{t},~ t\bar{t}H,~ t\bar{t}Z,~ t\bar{t}b\bar{b},~ t\bar{t}W, WH.
\label{eq:backgrounds}
\er
Among these background processes, $\rm t\bar{t}, t\bar{t}h, t\bar{t}Z, t\bar{t}b\bar{b}, t\bar{t}W$ contribute to both the hadronic and leptonic final state Eq.~\ref{eq:signal1},\ref{eq:signal2}, while $\rm WH$ becomes important only for the leptonic final state Eq.~\ref{eq:signal2}.

In practice, the impact of SMEFT operators on any SM process are studied 
considering only one operator at a time setting others to zero, 
and then repeat this exercise for rest of the operators. In fact,  
initially we followed this strategy to understand the 
effects of individual operators to tHq process, which can be confirmed also by looking at the FIM (Eq.~\ref{eq:fim}). However, in order to be realistic, effects of multiple relevant WCs are studied together to observe the combined effect, 
in particular due to the interference between 
different EFT operators and as well as with the SM. 
In our case, we set the values of the WCs to their 
respective best-fit as presented in Table \ref{tab:bestfit} (copied from second row of the Table~\ref{tab:quad}), which we 
dub as ``SM+EFT'', representing combined effects of the SM and  
the `EFT' operators (Eq~\ref{eq:smeft}). 

\begin{table}[H]
	\caption{\small {Bestfit values of the WCs, referred together as `SM+EFT'.}}
	\centering
	\begin{tabular}{ccccccc}
		\hline  
		\hline
		&  $\mathcal{O}_{t\phi}$ & $\mathcal{O}_{\phi Q}^{(3)}$ & $\mathcal{O}_{tW}$ & $\mathcal{O}_{Qq}^{(3,1)}$ &  $\mathcal{O}_{Qq}^{(3,8)}$& $\mathcal{O}_{\phi W}$ \\
		\hline
		Best-fit values&  -0.45$^{+3.8}_{-3.1}$ & -0.2$^{+3.4}_{-1.8}$ & -1.3$^{+2.6}_{-0.7}$ & -0.5$^{+1.3}_{-0.2}$ & -0.2$^{+0.5}_{-0.4}$&1.0$^{+0.7}_{-1.2}$ \\
		\hline
		\hline
	\end{tabular}
	\label{tab:bestfit}
\end{table} 

\subsection{Simulation of events and object selections}
As before, the events are generated in 
\nolinkurl{MG5aMC_atNLO-3.3.0}\cite{Alwall:2014hca} using 
\nolinkurl{SMEFTatNLO} UFO~\cite{Degrande:2020evl} package, which takes care of the 
SMEFT WCs. For decays, we use \nolinkurl{Madspin} taking into account of the 
SMEFT effects, which is crucial for the top-quark decay. 
The showering and hadronization are performed using 
\nolinkurl{PYTHIA8}\cite{Sjostrand:2006za,Sjostrand:2007gs} followed by 
detector simulation using Delphes~\cite{deFavereau:2013fsa} 
with the choice of CMS specific detector card. The selection of 
the two different kinematic regions (a) and (b)  are ensured by 
restricting Higgs boson $\rm p_T$ at the level of 
matrix element generation in \nolinkurl{Madgraph}. The hard-scattered jets are 
required to have $\rm p_T>20~GeV$, while choosing MLM 
matching with XQCut 20~\cite{Hirschi:2015iia}.

Apart from the signal events, this procedure is also followed for backgrounds with EFT effects, whereas, for SM-only backgrounds (i.e. all $C_i=0$), the common SM UFO model in \nolinkurl{Madgraph5-aMC@NLO} is used, instead of the \nolinkurl{SMEFTatNLO} one. In case of WH production, 
leptonic mode of W decay is considered, since this background contributes 
only to the leptonic final state. Similar to the signal events, background 
events are also generated in $\rm p_T$ bins, namely, 
$\rm 300<p_T<600~GeV$ (Bin 1) and $\rm p_T>600~GeV$ (Bin 2), 
of bosons (H, Z or W) in $\rm t\bar{t}h,~ t\bar{t}Z,~ t\bar{t}W$ events. 
For the generation of $\rm t\bar{t}$ and  $\rm t\bar{t}b\bar{b}$ events, 
same selections are imposed on the $\rm p_T$ of the top-quark.

All the objects are reconstructed using Delphes inputs. 
In the following, the reconstruction procedure for different 
objects are 
described separately for two different regions (a) and (b), whereas the lepton selection is same for both the regions. The simulation strategy including the reconstructions of objects are described below.

{\bf (a) Boosted-region:} To generate statistically 
appropriate number of events,  the boosted category is further divided in two separate Higgs $\rm p_T$ bins, $\rm 300<p_T(H)<600~GeV$ (Bin 1) and $\rm p_T(H)>600~GeV$ (Bin 2).

{\bf Higgs-jet(HJ):} First, e-flow objects of Delphes are used to construct fat-jets using Fastjet3.3.2\cite{Cacciari:2011ma} with Cambridge-Aachen\cite{Dokshitzer:1997in} algorithm setting jet size parameter $\rm R=1.0$. Minimum $\rm p_T$ of the fatjets is set to be 300 GeV and 600 GeV for Bin-1 and Bin-2 respectively. These fat jets are then passed through mass-drop Tagger (MDT)\cite{Butterworth:2008iy,Dasgupta:2013ihk} setting $\mu$ =0.667 and $\rm y_{cut} >0.09$ to remove contamination due to soft radiation. The subjets of the `tagged fat jet' are further matched with the b-quarks of the event which are selected within $|\eta|<2.5$ and with a matching cone $\rm \Delta R<0.3$. When both the subjets are found to be b-like, and mass of the fat-jet lies within the window $\rm m_{J}\in[100,150]$, we identify the tagged fatjet as the HJ. The presence of B-hadron in the b-like subjets are found to exist for about $95\%$ of the cases.

{\bf Top-jet(TopJ):} Hadronic Top-jets are constructed taking the 
Delphes objects which lie outside the HJ ($\rm \Delta R>R$) 
using Fastjet-3.3.2 with Cambridge-Aachen algorithm, with an
initial radius parameter $\rm R'=1.6$ and $\rm p_T>200 ~GeV$. 
The reconstructed fat-jets are passed through HEPToptagger~\cite{Plehn:2010st,Kasieczka:2015jma} to identify 
possible top-jets, with the choice of optimal-R requirement. 
When a fat-jet is tagged as top-jet, the optimized radius is noted as `$\rm R_{opt}$' for further use. It is to be noted that, for leptonic category, top-jet is not reconstructed due to the presence of neutrino leading to missing momentum.

{\bf Other-jets:} After HJ and TopJ reconstructions, Delphes objects 
which are away from HJ and TopJ (i.e, $\rm \Delta R>R,R_{opt}$)  
(for leptonic channel, distance from TopJ is not required), 
are clustered using Anti-kT algorithm with the radius parameter 
$\rm R''$=0.4 and $\rm p_T>20 ~GeV$. The event is accepted if there exists at least one jet.
On the other hand, for semi-leptonic final state, jets are matched with the 
b-quarks after the HJ tagging, 
to find out remaining b-jets of the event. These b-jets are part of the signal for the semi-leptonic case.

{\bf (b) Non-boosted region:} In the case of 
non-boosted region, obviously neither the Higgs nor the top quarks are tagged as fatjet. Instead, using e-flow objects of Delphes, we construct anti-kT jets of radius parameter $\rm R'''=0.4$ and minimum $\rm p_T$ cut of 20 GeV. Matching these jets with the b-quarks of the event using a matching cone $\rm \Delta R<0.3$, a set of b-jets are identified.

{\bf Higgs-jet(HJ):} In order to reconstruct the Higgs boson, the best possible combination of b-like jets are selected by minimizing $\rm (m_{b\bar{b}}-m_H)^2$, where $\rm m_H$ is set to 125 GeV. If $\rm m_{b\bar{b}}$ lies within the mass window $\rm m_H\in[100,150]$, then the corresponding combination of $\rm b\bar{b}$ system and the invariant mass is assumed to represent the Higgs boson.

{\bf Top-jet(TopJ):} Out of the remaining set of ordinary jets 
and the b-jets, the right combination is picked up, 
which minimizes $\rm (m_{(jjb)}-m_t)^2+(m_{(jj)}-m_W)^2$, and that represents a top-jet. For the leptonic case, top quark is not fully reconstructed. Instead, the lepton-b system originating from top quark decay is used as a substitute to construct top-quark related observables.

{\bf b-jets and other non-b jets:} After reconstruction of Higgs boson and top-quark (for hadronic channel) at least one jet (hadronic) or one b-jet and one ordinary jet (leptonic) are required.

{\bf Lepton selection :} Both leptons ($e$ and $\mu$) are selected with $\rm p_T>20~GeV$ and $|\eta|<2.5$.  The isolation of leptons are ensured using an isolation criteria taking into account the e-flow objects of Delphes as follows,
\br
\rm \frac{\sum p_{T}^{R<r}}{p_{T,\ell}}<I , \;\; \ell=e,\mu.
\er
Here the numerator presents the sum of $\rm p_T$ of all visible collections within a cone of radius R around the lepton, with $\rm r = \frac{10.0}{p_{T,\ell}}$ for boosted category, and $\rm r=0.3$ for non-boosted category, setting $\rm I =0.12$ and 0.25 for e and $\mu$ respectively. Notably, for the boosted category, a different isolation procedure with $\rm p_T$-dependent isolation cone is employed, familiar as mini-isolation \cite{CMS:2016krz}.

\subsection{Results}
\subsubsection{Effect of SMEFT operators to tHq process}
Several kinematic observables are constructed for each case (a) and (b) for ``SM+EFT", setting all WC to their best-fit values. We identify the right set of observables that show a clear deviation at the tail from the SM prediction. In order to quantify the sensitivities, we first estimate the level of contamination due to the SM backgrounds (Eq.~\ref{eq:backgrounds}) at the signal region, and then also checked the same including the EFT effects.

We demonstrate the distributions of a few interesting observables for the hadronic (Eq.~\ref{eq:signal1}) and leptonic final states (\ref{eq:signal2}), considering both the boosted and non-boosted regions.

\begin{figure}
	\begin{subfigure}[b]{0.5\textwidth}
		\centering
		\includegraphics[width=7.5 cm]{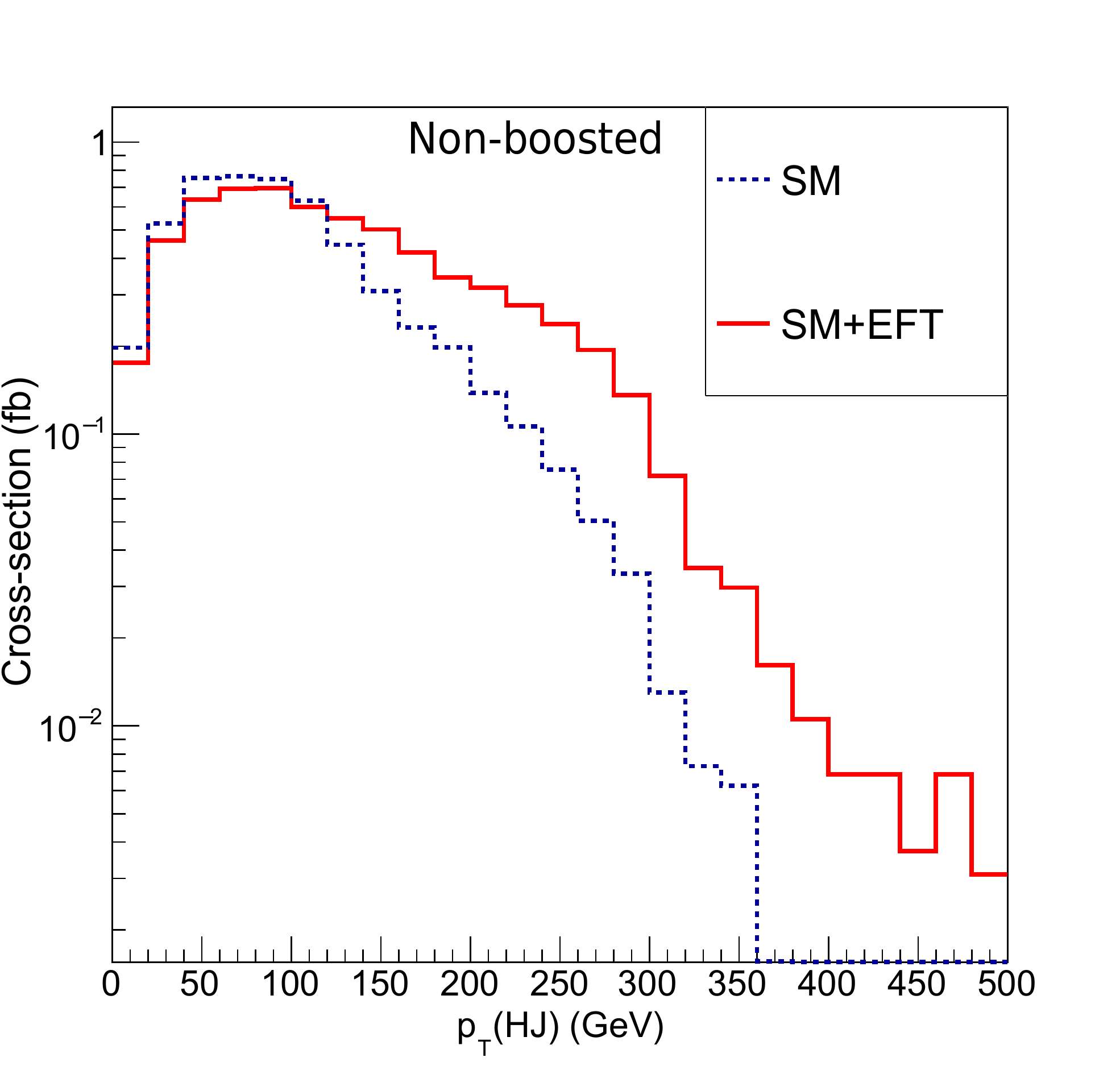}
	\end{subfigure}
	\begin{subfigure}[b]{0.5\textwidth}
		\centering
		\includegraphics[width=7.5 cm]{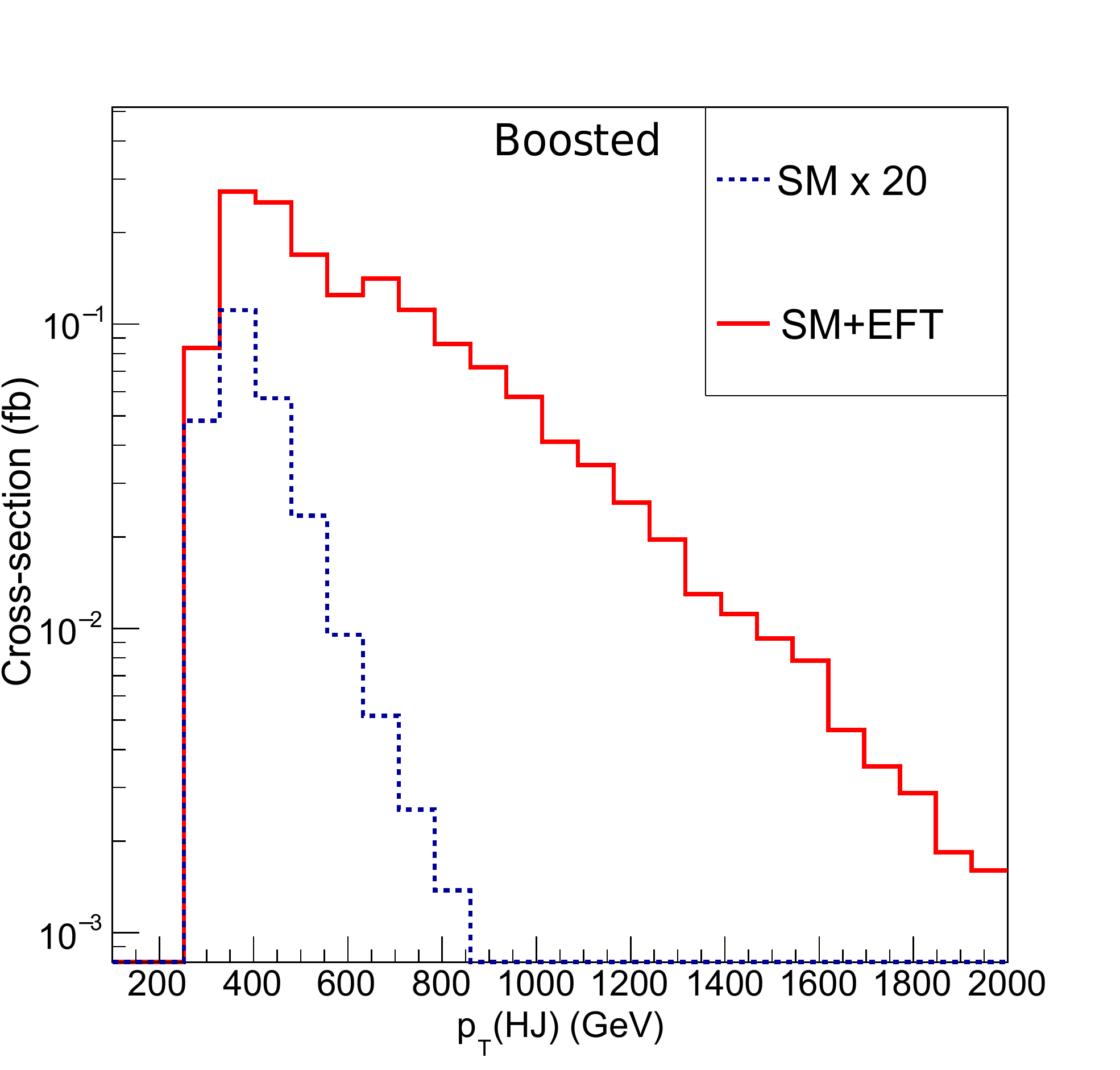}
	\end{subfigure}
	\caption{\small {Transverse momentum of the reconstructed Higgs boson (HJ) for non-boosted ($\rm p_T(H)<300~GeV$)(left) and boosted region ($\rm p_T(H)>300~GeV$)(right) for the hadronic final state.}}
	\label{fig:pth}
\end{figure}

{\bf $\bullet$ Hadronic final state:}
The cross-section normalized transverse momentum of the reconstructed Higgs boson for boosted (left) and non-boosted (right) cases are presented in Fig.~\ref{fig:pth}. The excess in the $\rm p_T$ spectrum is very large in the case of boosted regime, whereas it is less for the non-boosted case,
except beyond the $\rm p_T(H)>200~GeV$. The sudden fall, above $\rm p_T(H)>300~GeV$ is owing to a generator-level cut on the $\rm p_T$ of the Higgs boson. Notice that, for the boosted case,
the SM cross-section is multiplied by a factor of 20 for the sake of presentation.

\begin{figure}
	\begin{subfigure}[b]{0.5\textwidth}
		\centering
		\includegraphics[width=7.5 cm]{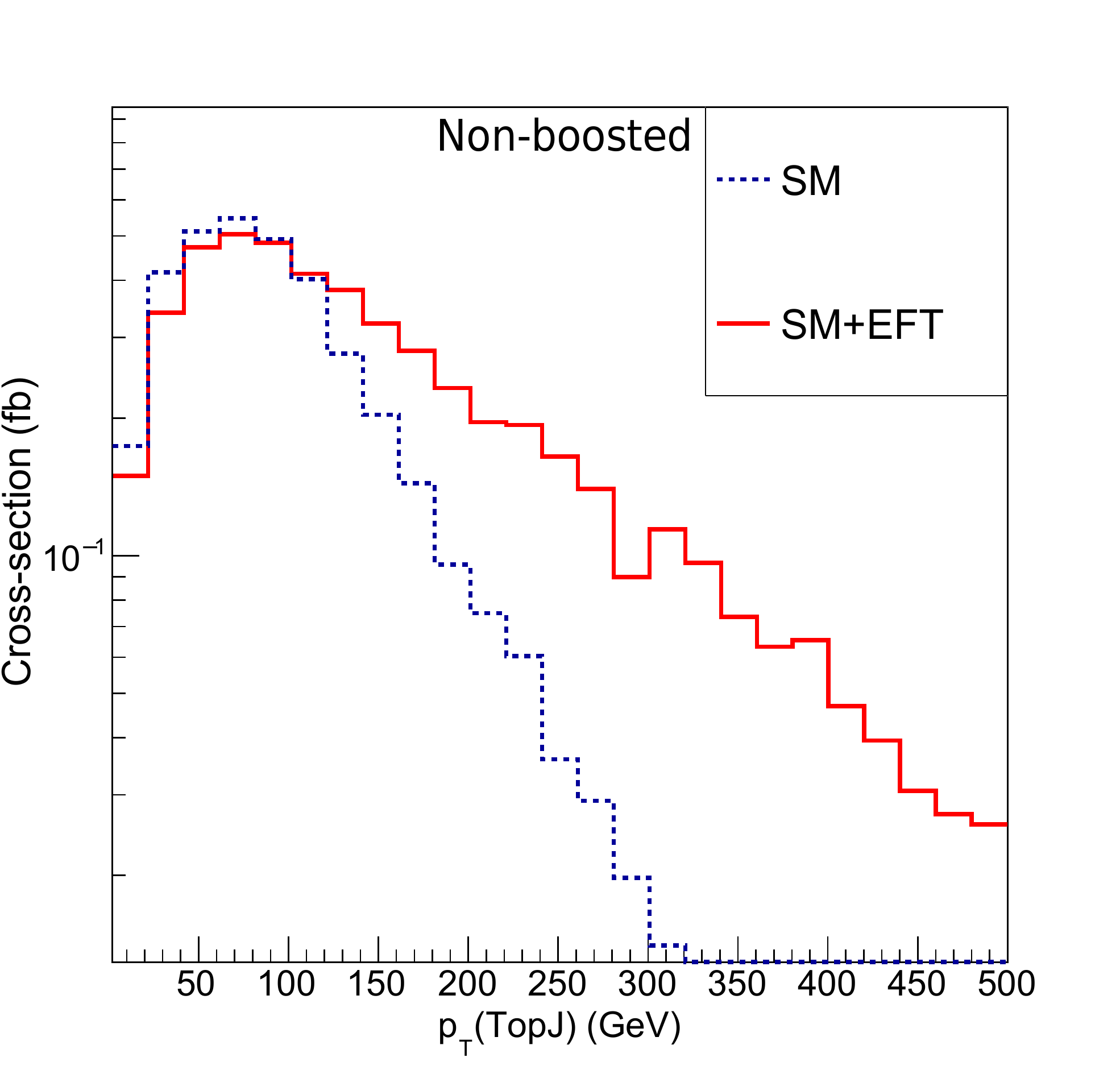}
	\end{subfigure}
	\begin{subfigure}[b]{0.5\textwidth}
		\centering
		\includegraphics[width=7.5 cm]{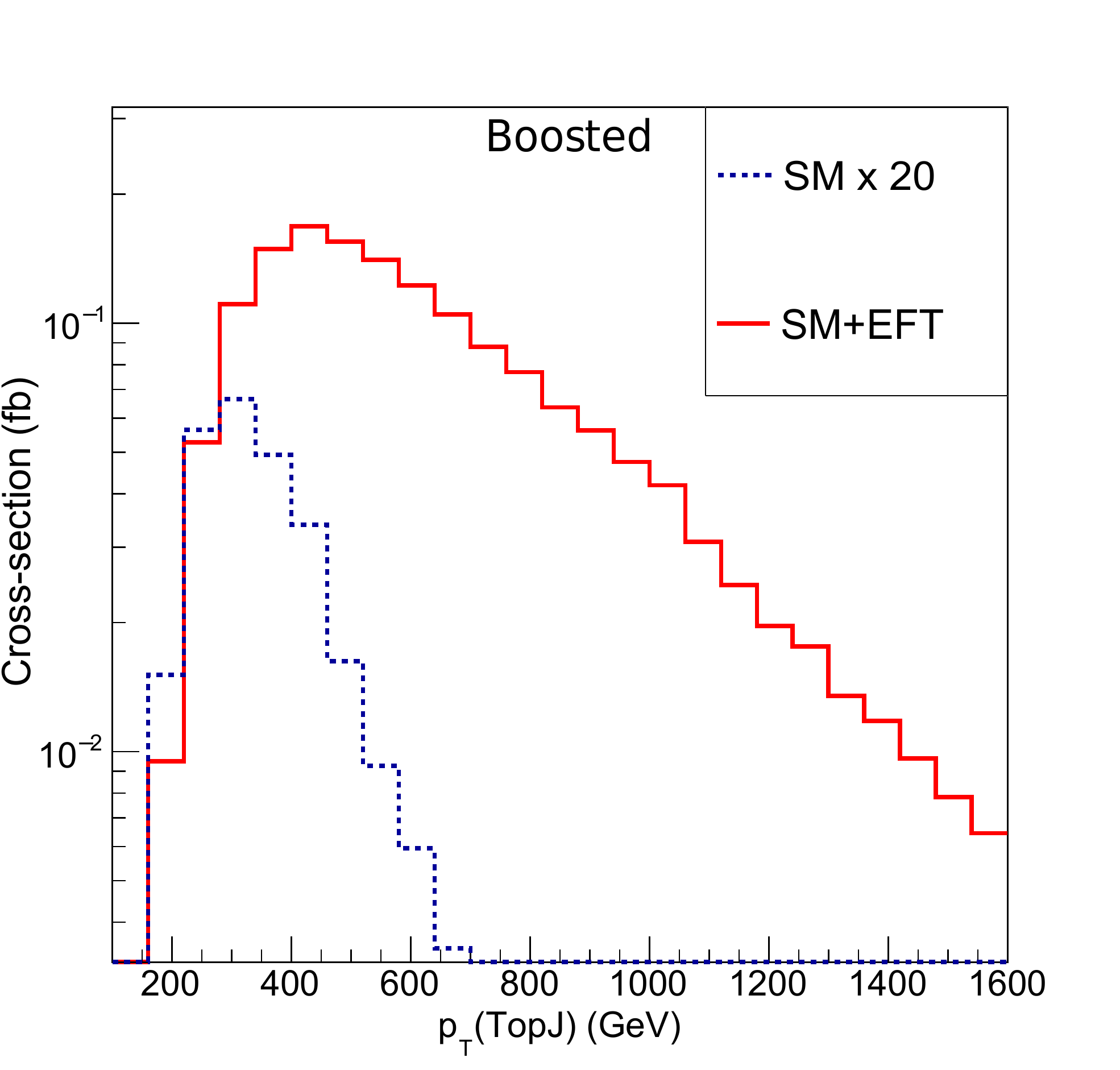}
	\end{subfigure}
	\caption{\small {Transverse momentum of the reconstructed Top-jet (TopJ) for the hadronic final state in the same set-up as Fig.~\ref{fig:pth}.}}
	\label{fig:pttop}
\end{figure}

Similarly, the distribution of $\rm p_T$ of the top-jet is shown in Fig.~\ref{fig:pttop} both for SM and SM+EFT cases. As no generator-level selection is imposed on the $\rm p_T$ of the top-quark or the light hard-scattered quark, the distribution in the non-boosted region does not show any abrupt fall as seen before at the higher side of the HJ $\rm p_T$ distribution. In this distribution also, a significant deviation is observed on the higher side, in particular for the boosted case.

\begin{figure}
	\begin{subfigure}[b]{0.5\textwidth}
		\centering
		\includegraphics[width=7.5 cm]{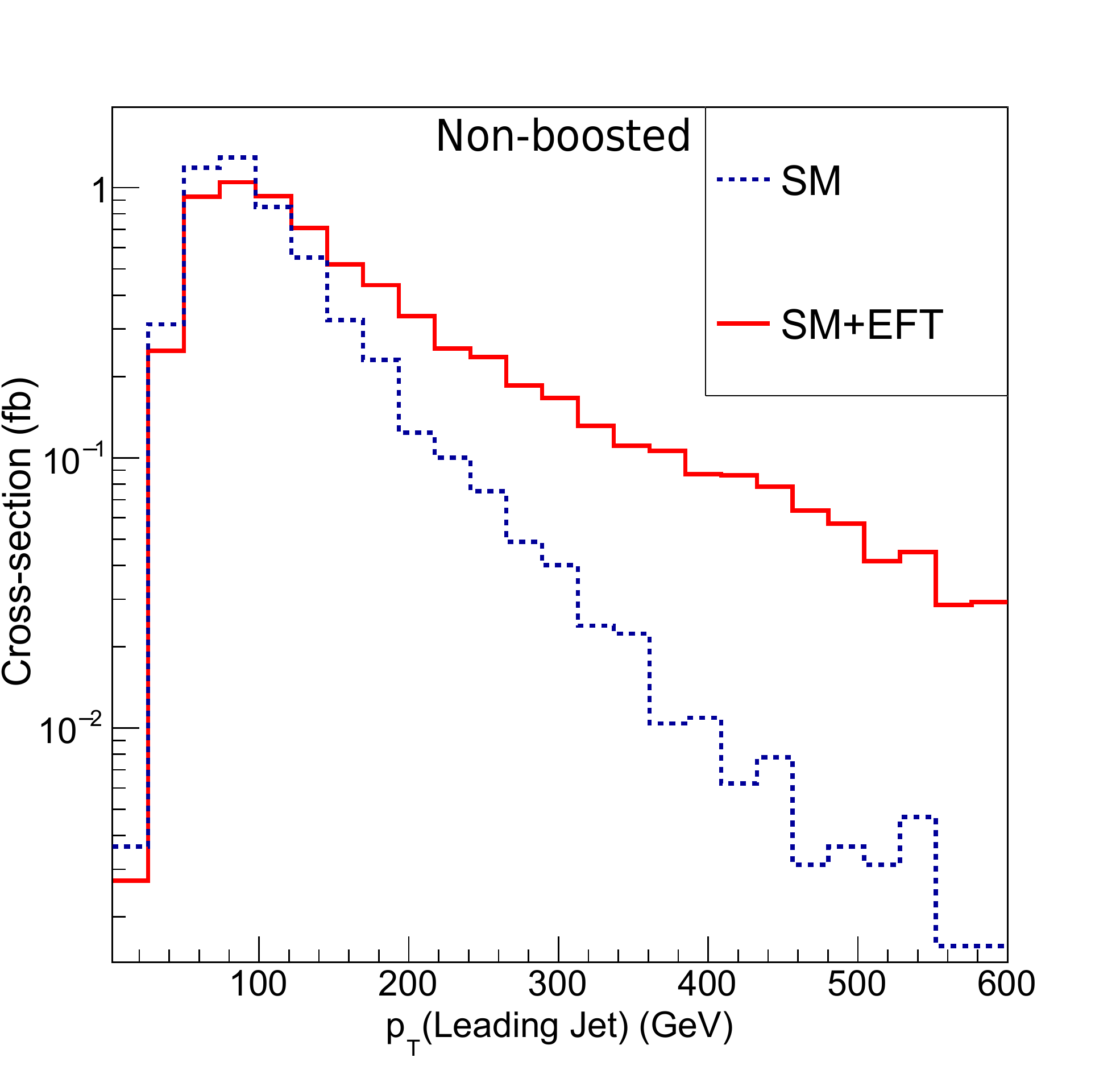}
	\end{subfigure}
	\begin{subfigure}[b]{0.5\textwidth}
		\centering
		\includegraphics[width=7.5 cm]{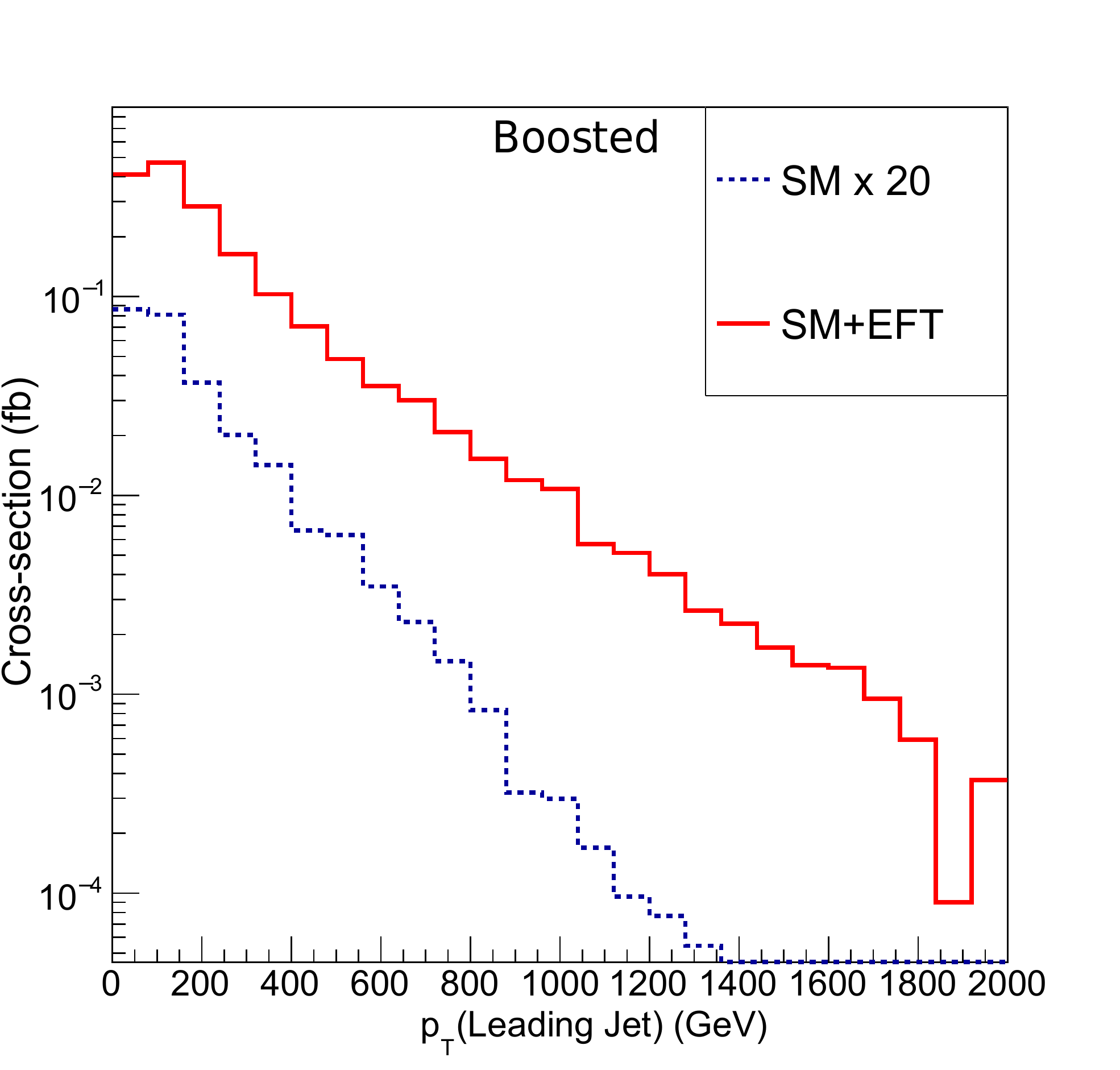}
	\end{subfigure}
	\caption{\small {Transverse momentum of the reconstructed leading jet for the hadronic final state in the same set-up as Fig.~\ref{fig:pth}.}}
	\label{fig:ptjet}
\end{figure}

\begin{figure}
	\begin{subfigure}[b]{0.5\textwidth}
		\centering
		\includegraphics[width=7.5 cm]{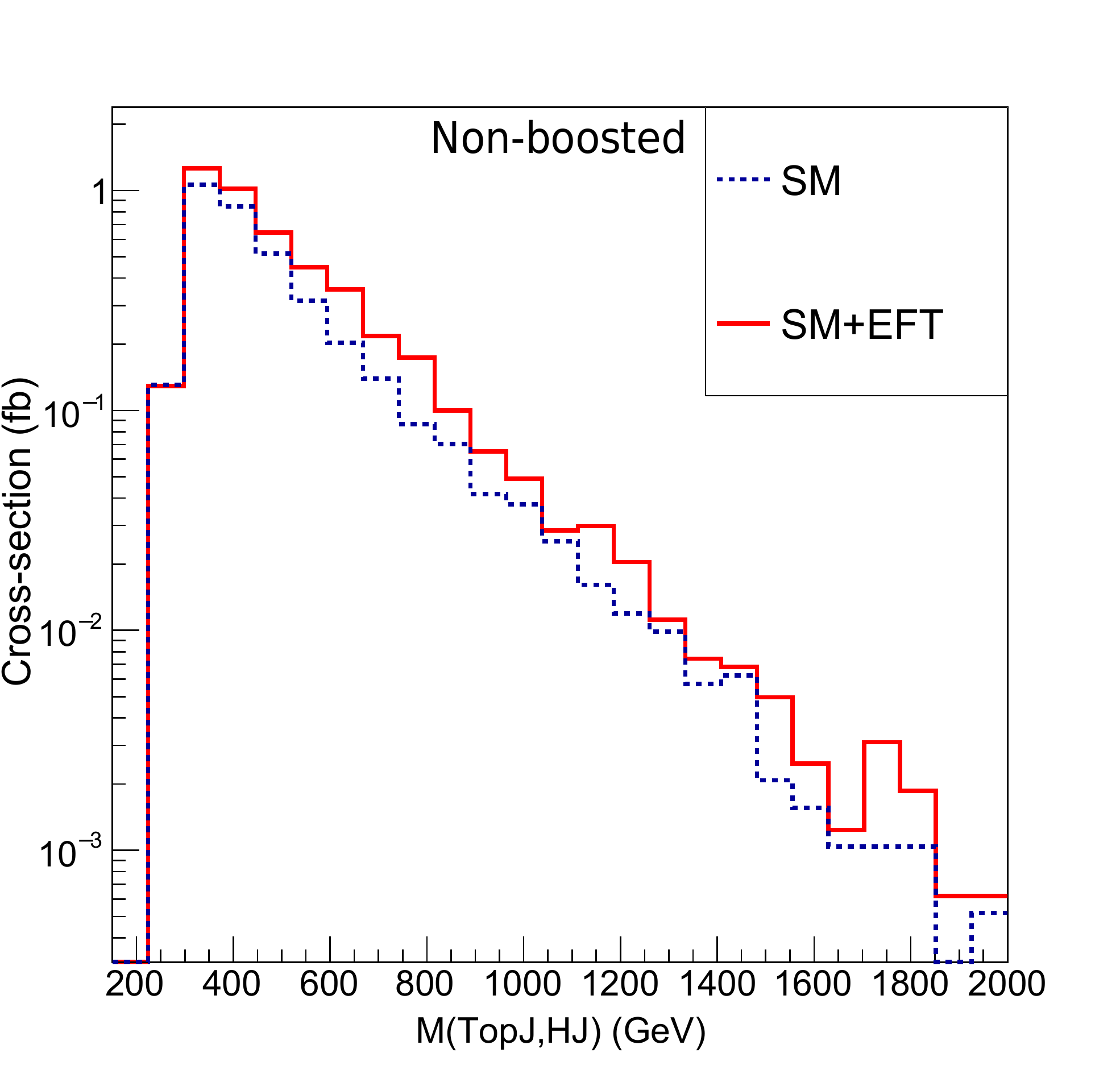}
	\end{subfigure}
	\begin{subfigure}[b]{0.5\textwidth}
		\centering
		\includegraphics[width=7.5 cm]{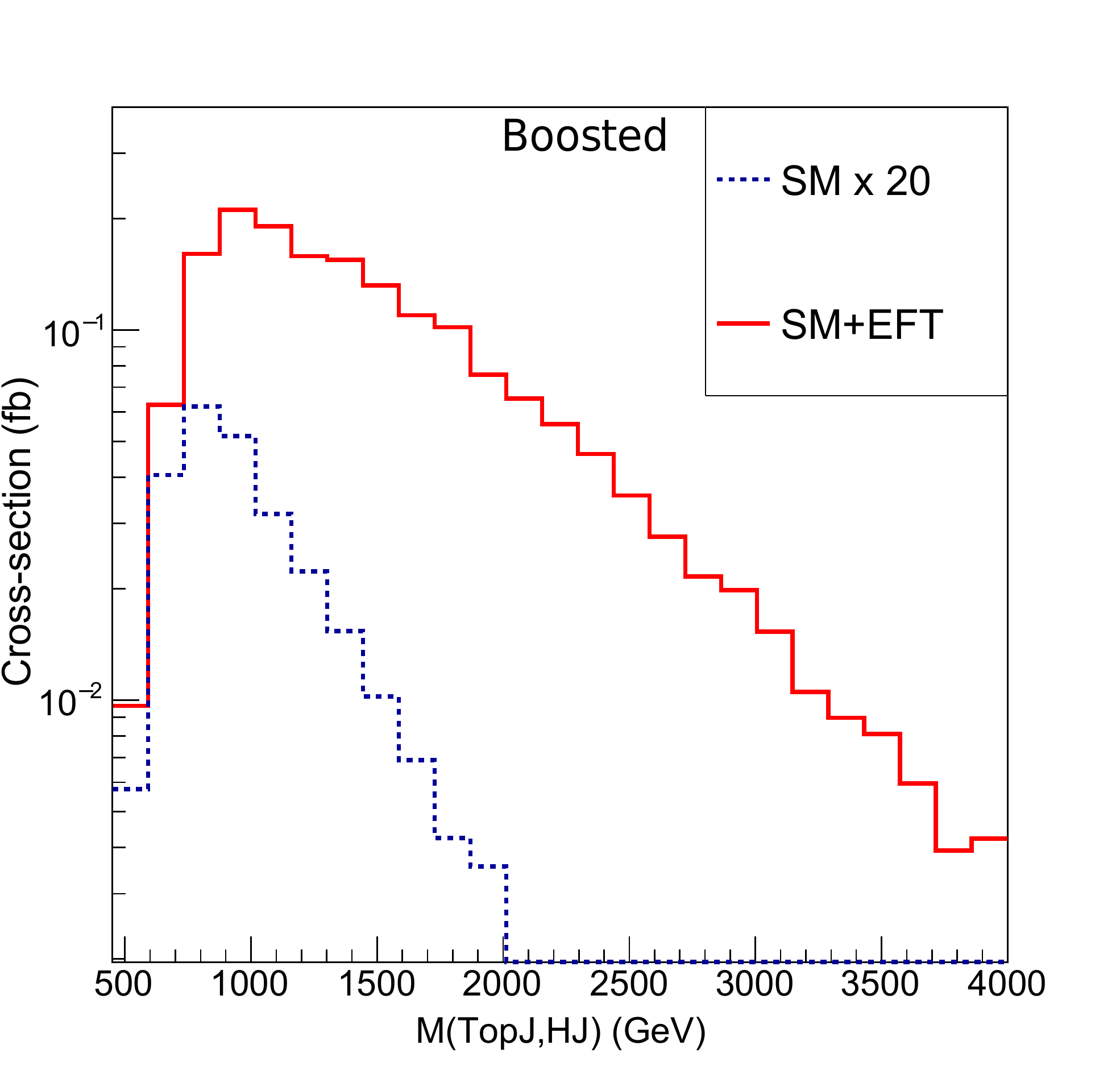}
	\end{subfigure}
	\caption{\small {Invariant mass of the Higgs-Jet and Top-jet for the hadronic final state in the same set-up a Fig.~\ref{fig:pth}.}}
	\label{fig:invmth}
\end{figure}

The $\rm p_T$ of the leading-jet, as presented in Fig.~\ref{fig:ptjet}, depicts a mild deviation at the tail, unlike the case of HJ and TopJ. It can be attributed to the fact that the vertex from where the jet originates are not affected by any energy growth due to the scattering sub-amplitude $\rm bW\to tH$ ~\cite{Farina:2012xp}.
The deviations of the TopJ and HJ at the tail of the distributions are also expected to be reflected in the invariant mass of these two objects, as seen in Fig.~\ref{fig:invmth}. Remarkably, a very clear deviation is observed, in particular for the boosted case.

\begin{figure}
	\begin{subfigure}[b]{0.5\textwidth}
		\centering
		\includegraphics[width=7.5 cm]{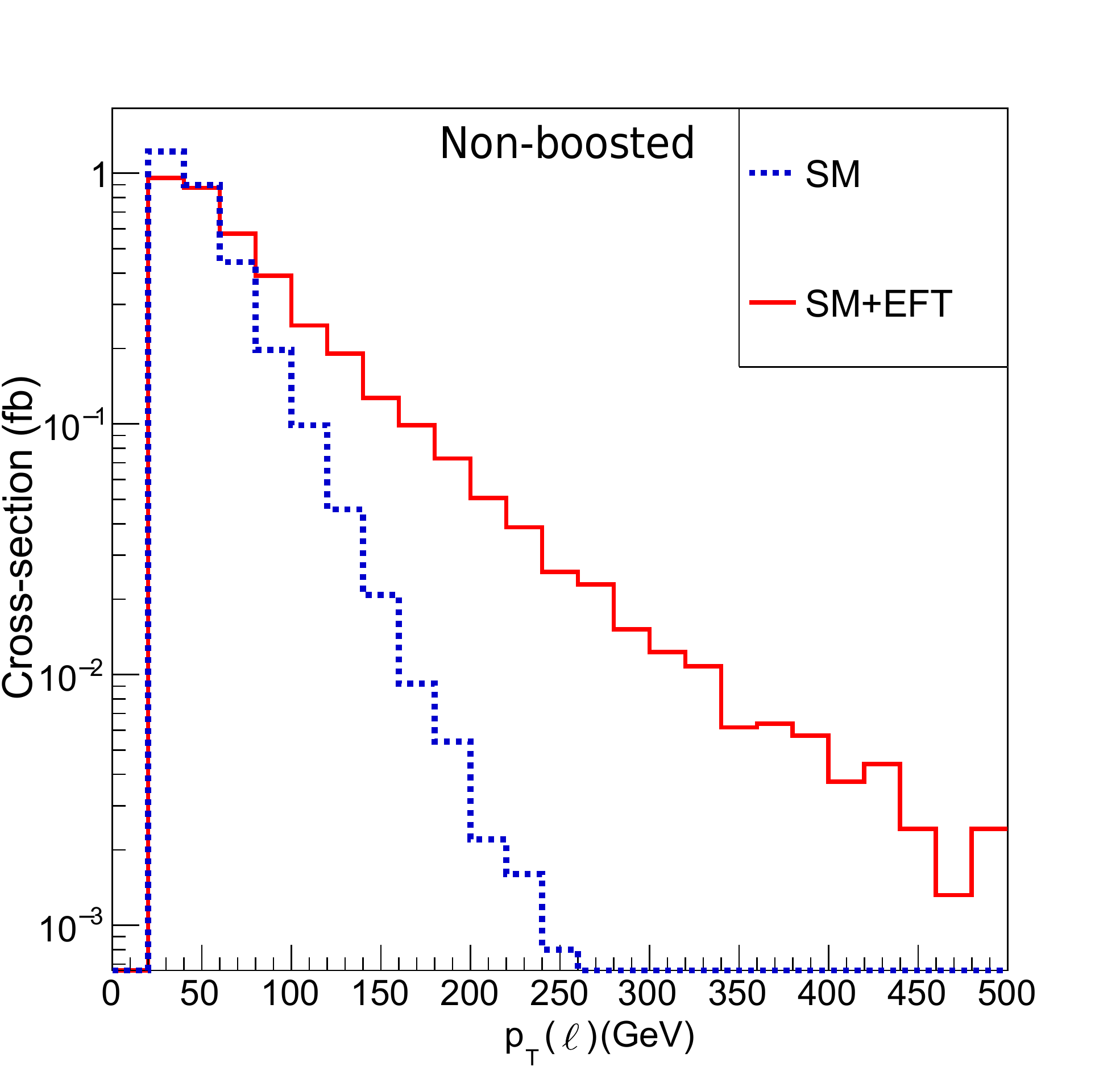}
	\end{subfigure}
	\begin{subfigure}[b]{0.5\textwidth}
		\centering
		\includegraphics[width=7.5 cm]{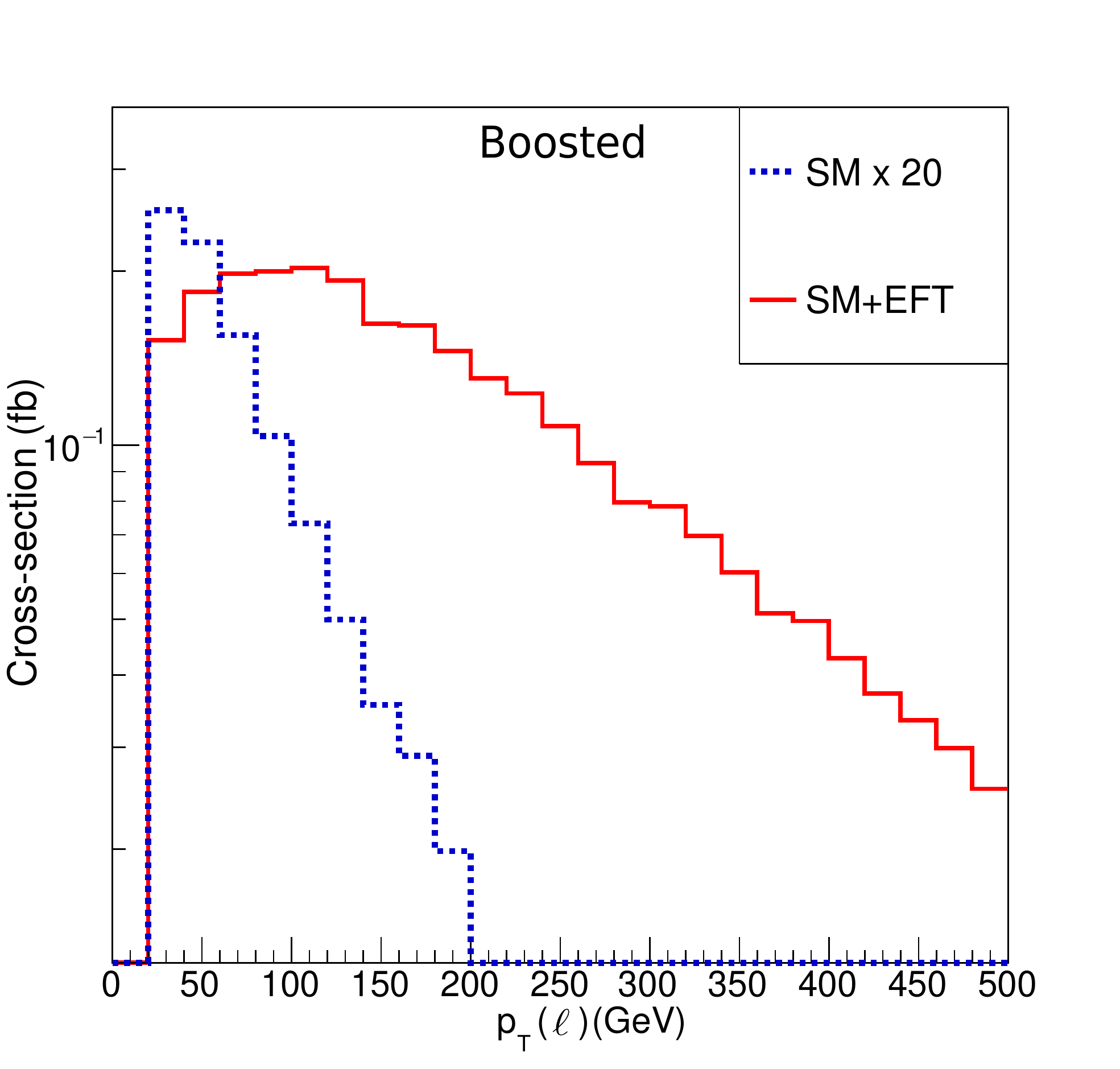}
	\end{subfigure}
	\caption{\small {Transverse momentum of the isolated lepton for the leptonic final state in the same set-up as Fig.~\ref{fig:pth}.}}
	\label{fig:ptlep}
\end{figure}

{\bf $\bullet$ Leptonic final state:}
The distribution of $\rm p_T$ of the lepton arising from the decay of top-quark is presented in Fig.~\ref{fig:ptlep}. As expected, the lepton receives the similar feature of the $\rm p_T$ of the top, and exhibits substantial deviation as the TopJ. In this case, particularly for the boosted category, the effect is much more pronounced than the $\rm p_T$ of TopJ for hadronic case (Fig.~\ref{fig:pttop}).

\begin{figure}
	\begin{subfigure}[b]{0.5\textwidth}
		\centering
		\includegraphics[width=7.5 cm]{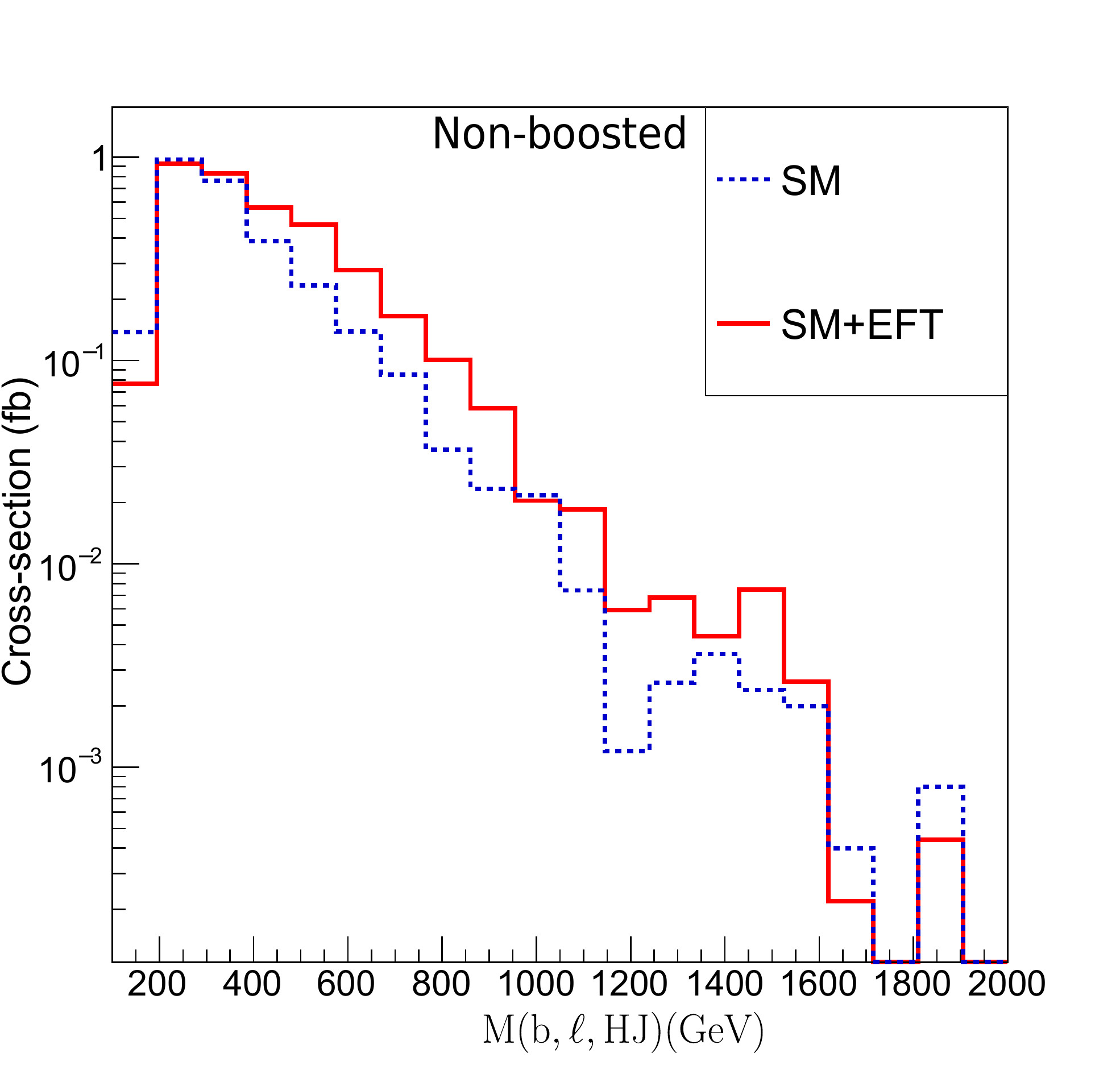}
	\end{subfigure}
	\begin{subfigure}[b]{0.5\textwidth}
		\centering
		\includegraphics[width=7.5 cm]{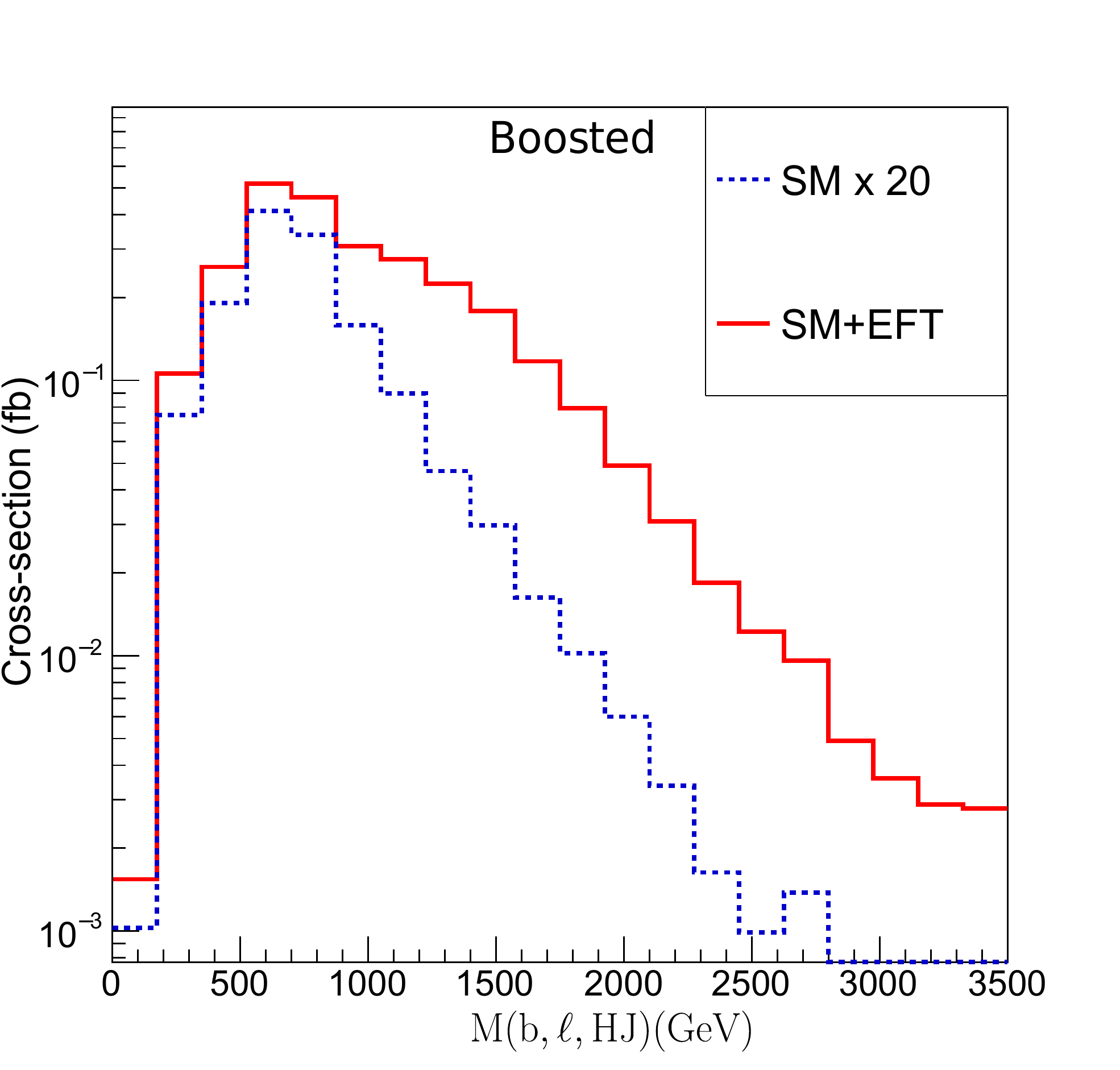}
	\end{subfigure}
	\caption{\small {Invariant mass of the lepton, b-jet and Higgs-jet system for the leptonic final state in the same set-up as Fig.~\ref{fig:pth}, requiring $\rm \Delta R (\ell,b)<1.5$.}}
	\label{fig:invmhlb}
\end{figure}

Equivalent to the M(TopJ,HJ) observable for the hadronic case, the invariant mass of the lepton, HJ, and the b-jet is constructed. The b-jets associated with the lepton are identified by requiring the criteria $\rm \Delta R(\ell,b)<1.5$. In Fig.~\ref{fig:invmhlb}, the invariant mass of $\rm b\mhyphen \ell \mhyphen HJ$ system is presented for the non-boosted (left) and boosted (right) case. Remarkable deviations are observed at the high value of the invariant mass for the boosted case.

\subsubsection{Signal and backgrounds}
The visible effects of the relevant SMEFT operators in various 
kinematical observables for the tHq process are presented in the previous section. Deviations from the SM prediction is observed at the tail of the distributions. Discovery potential of the signatures of these operators can be realized once we make a proper estimation of background contamination in the signal regions. The possible dominant sources of SM backgrounds 
are presented in Eq.~\ref{eq:backgrounds}. Now we proceed to find the signal sensitivity, by estimating signal significance for the boosted region of the tHq process.

For the hadronic final state, QCD multi-jet events, having huge cross-section, may add fake contribution as one of the backgrounds. We have checked simulating a sufficient number of QCD events that the contribution is small, though non-negligible compared to the signal in the boosted region with only the minimal selection criteria for the final state Eq.~\ref{eq:signal1},and \ref{eq:signal2}. However, after implementing other discriminating variables, the QCD contribution is expected to be reduced further and even a multivariate analysis (MVA) may help to bring it down substantially. Due to the lack of appropriate statistics of events after the basic selections, further study turned out to be beyond our scope, and we conclude that the proper estimation of the QCD background is to be done experimentally by a data-driven analysis.

The cross-section of each of the background processes at the leading 
order (LO) are presented in Table~\ref{tab:yields_hadronic} 
and \ref{tab:yields_leptonic}, subject to different set of selection cuts as required. For the sake of comparison, 
the cross-section including EFT (`SM+EFT') in the calculation are also presented.
Moreover, it is important to note that the SM background processes are simulated including 
the effect of EFT operators like the signal process, 
setting the respective WCs to the best-fit values as presented 
in Table \ref{tab:bestfit}. 
For the background processes, the numbers in the parenthesis  
represent the cross-section yields due to  
SM-only contribution, i.e., setting all WCs are zero. 
The first row presents the fiducial production cross-section 
at leading order (LO), i.e., subject to the generator-level phase 
space selections described above, while later rows show 
the cross-section yields after imposing respective selection cuts.

\begin{table}
	\caption{\small {Signal and background cross-section yields (in fb) for the hadronic final state after basic selections including EFT effects in both cases. The numbers in the parenthesis represent the SM-only background contributions.}}
	\centering
	\begin{adjustbox}{max width=\textwidth} 
		\begin{tabular}{ccccccc}
			\hline  
			\hline
			& SM+EFT & $\rm t\bar{t}H(b\bar{b})$ & $\rm t\bar{t}b\bar{b}$ & $\rm t\bar{t}$ & $\rm t\bar{t}Z$ & $\rm t\bar{t}W$ \\
			\hline
			Cross-section(fb) (LO)& 21  &21(17)  & 59(59)& 15410(15300) & 39(39)&13(13)\\
			No. of HJ =1  &9.6 & 8.5(6.4) & 1.3(1.1) & 27.8(23.9) & 0.7(0.7)&0.08(0.09)\\
			$\rm No.~ of~ TopJ=1$  & 1.9 & 1.3(0.9) & 0.2(0.16) &  3.0(2.9) & 0.08(0.07)&0.005(0.002)\\
			$\rm No.~ of~ Jets\geq 1$  & 1.6 & 1.2(0.9) & 0.2(0.15) &  2.6(2.5) & 0.08(0.06)&0.004(0.001)\\
			$\rm p_T(topJ)>400~GeV$  & 1.3 & 0.6(0.4) & 0.1(0.09) &  1.3(1.25) & 0.03(0.02)&0.001(0.0008)\\
			$\rm M(topJ,HJ)>1000~GeV$  & 1.1 & 0.4(0.2) & 0.06(0.04) &  0.8(0.77) & 0.02(0.02)&0.0005(0.0003)\\
			\hline
			\hline
		\end{tabular}
	\end{adjustbox}
	\label{tab:yields_hadronic}
\end{table}

\begin{table}
	\caption{\small {Same as in Table~\ref{tab:yields_hadronic}, but for leptonic final state.}}
	\centering
	\begin{adjustbox}{max width=\textwidth} 
		\begin{tabular}{cccccccc}
			\hline  
			\hline
			& SM+EFT & $\rm t\bar{t}H(b\bar{b})$ & $\rm t\bar{t}b\bar{b}$ & $\rm t\bar{t}$ & $\rm t\bar{t}Z$ & $\rm t\bar{t}W$  & $\rm W(\ell\nu)H(b\bar{b})$\\
			\hline
			Cross-section(fb) (LO)& 7  &22(17)  & 59(59)& 15480(15300) & 39(39)&13(13) & 22.2(6.5)\\
			No. of HJ =1  	& 3.5 & 7.6(6.2) & 1.3(1.0) & 26.1(23.9) & 0.7(0.7) & 0.08(0.09) & 10.1(2.0)\\
			$\rm No.~ of~ \ell=1$  			& 2.8 & 2.2(1.8) & 0.4(0.3) &  6.5(6.2) & 0.2(0.2) & 0.02(0.03) & 8.1(1.6)\\
			$\rm No.~ of~ Jets\geq0$  				& 2.8 & 2.2(1.8) & 0.4(0.3) &  6.5(6.2) & 0.2(0.2) & 0.02(0.03) & 8.1(1.6)\\
			$\rm p_T(\ell)>300~GeV$  			& 0.7 & 0.09(0.06) & 0.03(0.01) &  0.5(0.3) & 0.004(0.002)& 0.003(0.002) & 1.6(0.1)\\
			$\rm M(\ell,b,HJ)>800~GeV$  		& 0.5 & 0.04(0.03) & 0.006(0.004) &  0.1(0.08) & 0.0003(0.0002)& 0.0004(0.0002) & 0.8(0.03)\\
			\hline
			\hline
		\end{tabular}
	\end{adjustbox}
	\label{tab:yields_leptonic}
\end{table}

The reconstruction of one Higgs jet (HJ) in the event is found to be very effective in eliminating an enormous amount of backgrounds, except for $\rm t\bar{t}H$, which includes a true Higgs boson in the final state like the signal. The presence of a single TopJ in the signal also helps in suppressing a substantial fraction of backgrounds owing to the higher top-reconstruction efficiency for the signal, since the population of high $\rm p_{T}$ top quarks in the signal process is higher compared to the backgrounds. Notably, demanding at least one QCD jet is important to confirm the signal events, although not very effective in reducing the background events. Finally, in order to eliminate further the remaining background yields, a cut on the $\rm p_T$ of the Top-jet and an invariant mass cut of the Top-Higgs system for the hadronic category are imposed. These two cuts essentially isolate the signal region where a substantial excess is visible, in particular, for the boosted case. These two selection cuts reject around 70$\%$ or more background events.
Similarly, for the leptonic final state, cut on the $\rm p_T$ of lepton followed by a requirement on the invariant mass of the lepton, b, and the HJ system, $\rm M(\ell,b,HJ)>800~GeV$, eliminate around $99\%$ of backgrounds, without costing too much signal events. The overall acceptance efficiency for the signal is about $5-8\%$ for both the hadronic and leptonic categories, whereas, for backgrounds, it is $\sim 0.001\% - 0.006\%$. Finally, the total background cross-section turns out to be $\sim 1.3$ fb and $\sim 0.9$ fb for hadronic and leptonic cases respectively.

\begin{figure}
	\begin{subfigure}[b]{0.5\textwidth}
		\centering
		\includegraphics[width=7.5 cm]{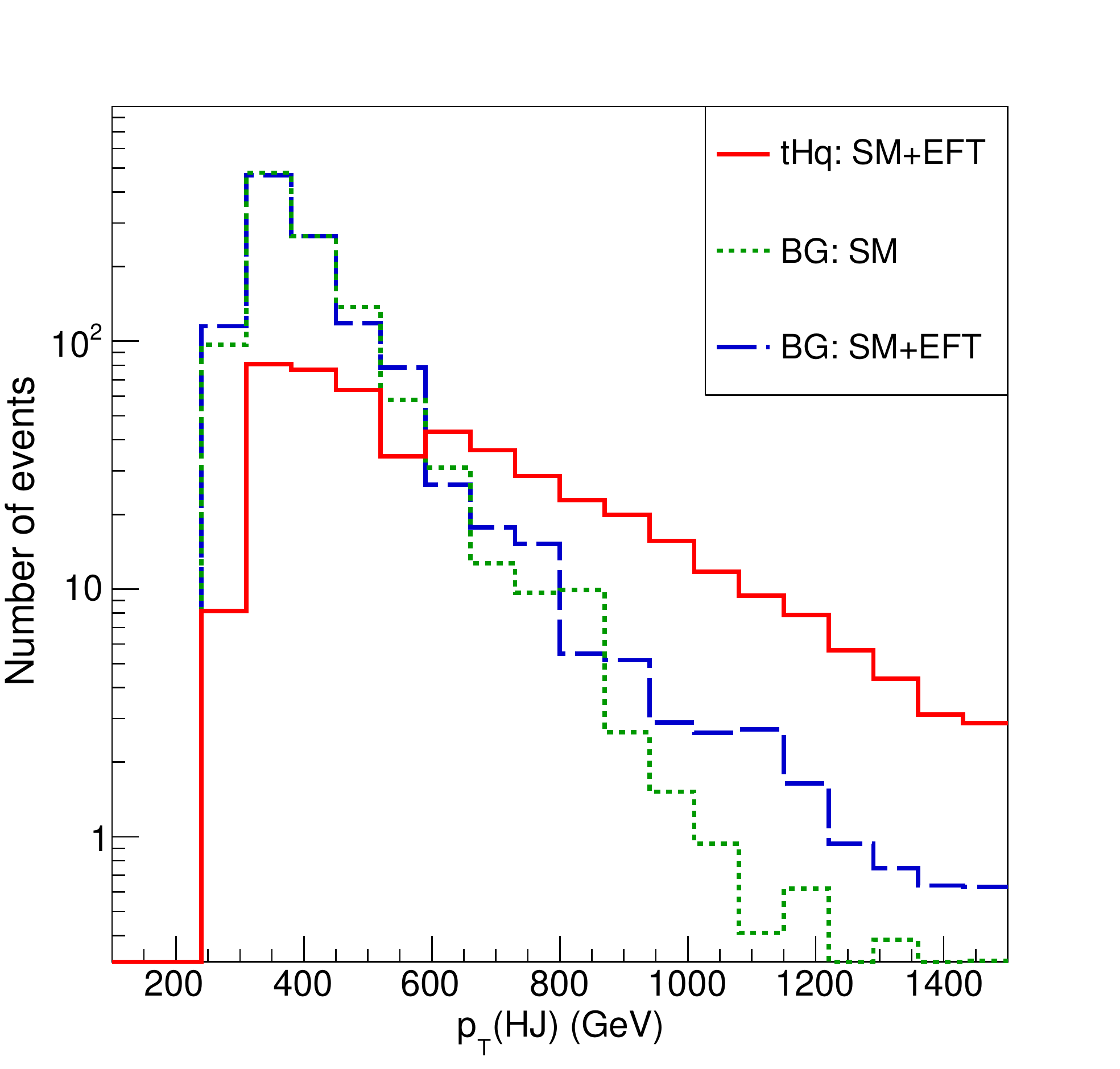}
	\end{subfigure}
	\begin{subfigure}[b]{0.5\textwidth}
		\centering
		\includegraphics[width=7.5 cm]{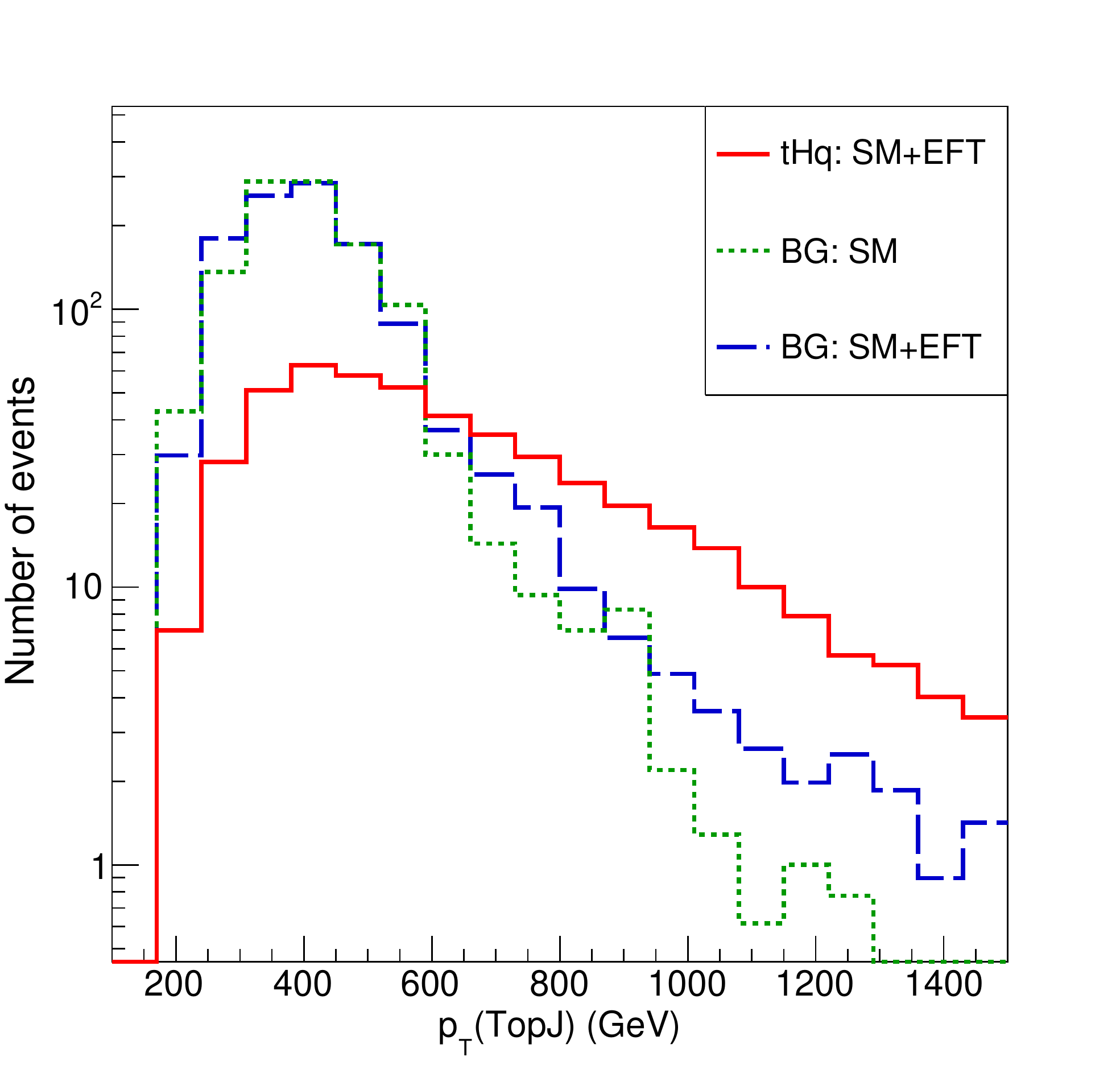}
	\end{subfigure}
	\caption{\small {Number of events for $\rm \mathcal{L}=300~fb^{-1}$ in $\rm p_T$-bins of Higgs-Jet (left) and Top-jet (right) in the boosted region ($\rm p_T(H)>300~GeV$) of the hadronic final state along with the total main backgrounds. `BG' corresponds to backgrounds, which can either be SM-only or with EFT effects (SM+EFT).}}
	\label{fig:bkg1}
\end{figure}

\begin{figure}
	\begin{subfigure}[b]{0.5\textwidth}
		\centering
		\includegraphics[width=7.5 cm]{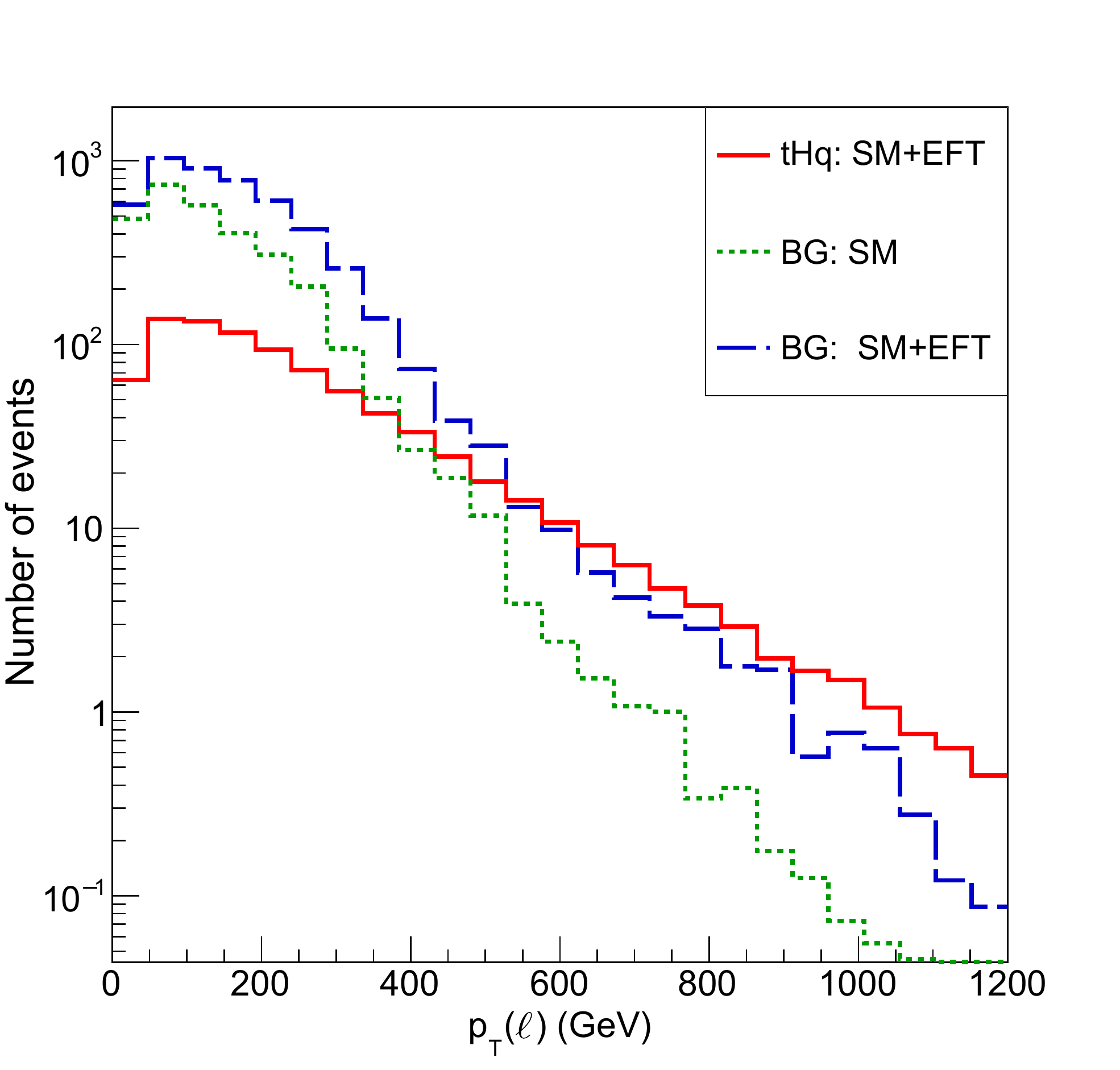}
	\end{subfigure}
	\begin{subfigure}[b]{0.5\textwidth}
		\centering
		\includegraphics[width=7.5 cm]{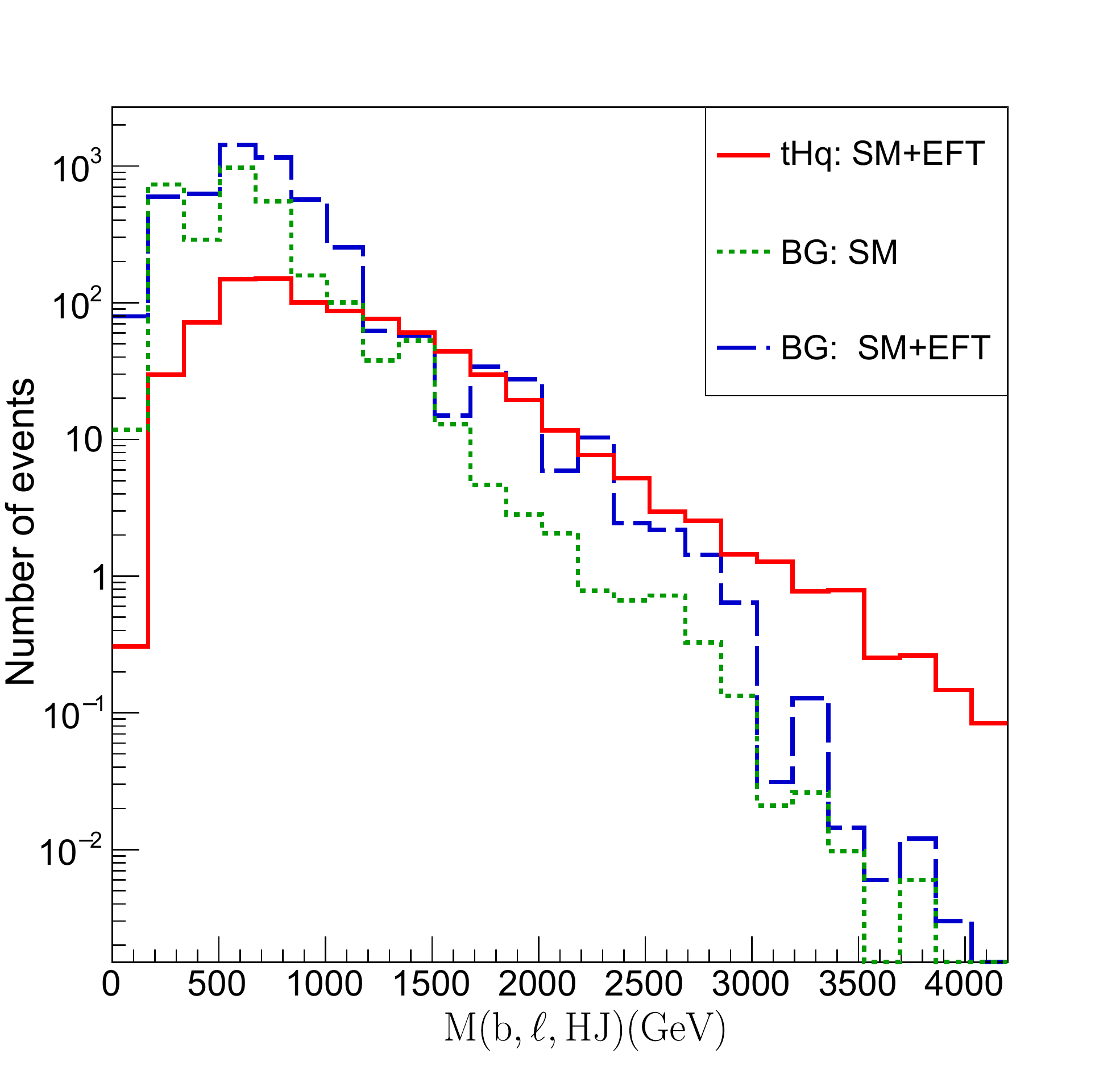}
	\end{subfigure}
	\caption{\small {Same as Fig.~\ref{fig:bkg1}, but in $\rm p_T$-bins of lepton (left) and invariant mass of Higgs-lepton-bjet system (right) for the leptonic final state.}}
	\label{fig:bkg2}
\end{figure}

The Fig.~\ref{fig:bkg1} presents the $\rm p_T$ distribution, normalized to luminosity
$\rm \mathcal{L}=300~fb^{-1}$, of
TopJ and HJ after all selection cuts for hadronic events. It also shows the clear excess of signal events
at the tail. Similarly, for the leptonic category, the distribution of $\rm p_T$ of lepton and invariant mass of lepton, b-jet, and reconstructed Higgs boson are presented in Fig.~\ref{fig:bkg2}.

The distributions in Fig.~\ref{fig:bkg1} and \ref{fig:bkg2} demonstrate that in the boosted region of the tHq production process, a countable excess of events can be observed for 
a luminosity $\rm \mathcal{L}=300~fb^{-1}$, which is expected to increase by 10 folds 
for $\rm \mathcal{L}=3000~fb^{-1}$. Alternatively, non-observation of any excess certainly can impose much stronger bounds on the WCs of the relevant operators.

\begin{figure}
	\begin{subfigure}[b]{0.5\textwidth}
		\centering
		\includegraphics[width=7.8 cm]{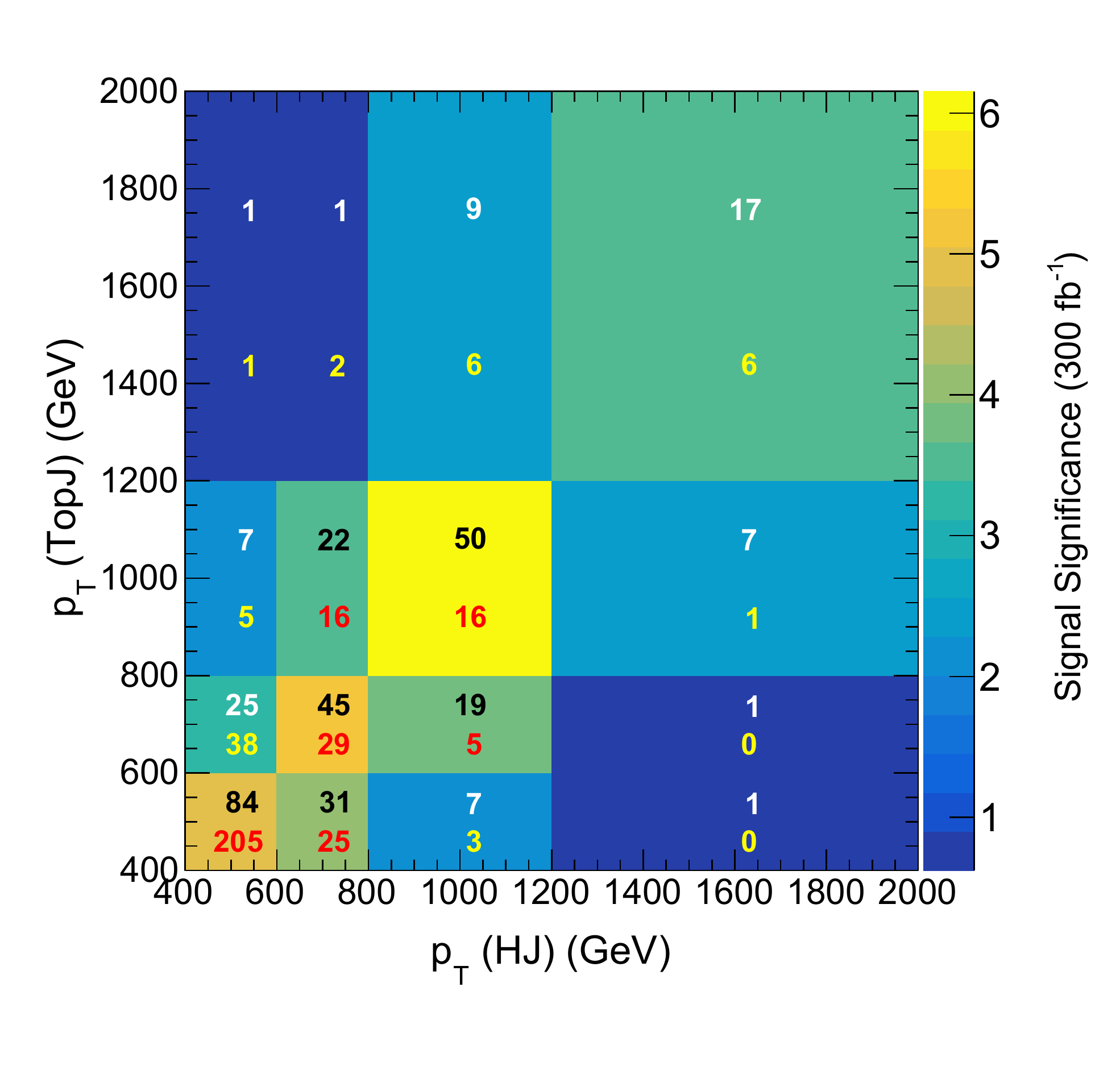}
	\end{subfigure}
	\begin{subfigure}[b]{0.5\textwidth}
		\centering
		\includegraphics[width=7.8 cm]{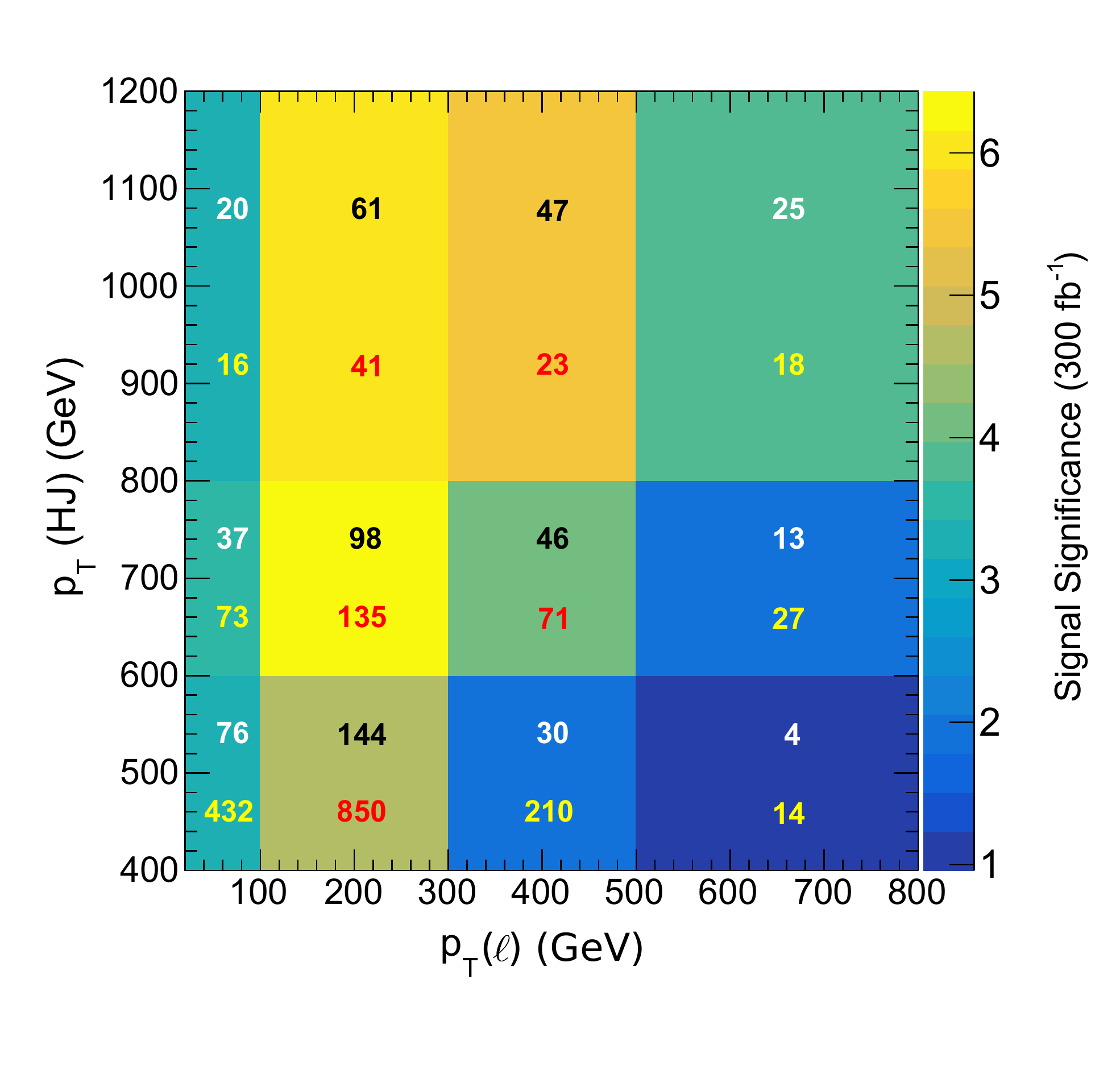}
	\end{subfigure}
	\caption{\small {2D distribution of $\rm p_T(HJ)$ and $\rm p_T(TopJ)$ or $\rm p_T(\ell)$ in terms of signal significance ($\rm S/\sqrt{(S+B)}$) at $\rm \mathcal{L}=300~fb^{-1}$ for hadronic (left) and leptonic (right) final state. The upper (lower) entries present the number of event yields for signal (total background).}}
	\label{fig:sens_2d}
\end{figure}

The feature of the `high scale sensitivity' reflected in the excess can be 
captured in the `bin by bin' estimation of the respective kinematic variable, rather than
counting a total number of events. Basically, the region where excess is observed is divided into several bins, and the number of events corresponding to those for both the signal and total background is counted. Since $\rm p_T$ of constructed kinematic variables as shown above are expected to be correlated, bins in the 2D plane are considered to be useful in presenting significances. In Fig.~\ref{fig:sens_2d} we present the signal region in 2D planes dividing 
into the $\rm p_T$ bins of the HJ and TopJ for hadronic(left) and leptonic(right)
case. In each bin, the upper (lower) entries present the signal (total background)
yields. The respective colors of each bin show the level of significances,
$\rm S/\sqrt{(S+B)}$, for a luminosity option $\rm \mathcal{L}=300~fb^{-1}$.

Notice that for the hadronic case, more than $3\sigma$ significance can be achieved for the $\rm p_T$ range 400-1200 GeV of both HJ and TopJ, whereas for the leptonic case, the range appears to be 20-800 GeV for lepton and 400-1200 GeV for HJ. The drop of the signal significance, assuming of $10\%$ background uncertainty, differs from bin to bin with a maximum drop of $\sim 30-50\%$ in the lower $\rm p_T$ bins, whereas a negligible reduction ($\sim 0$) towards the higher $\rm p_T$ region. The sensitivity may go up by almost a factor 3 for $\rm \mathcal{L}=3000~fb^{-1}$. Evidently, Fig.~\ref{fig:sens_2d} demonstrates the discovery potential of the signature of the impact of EFT in tHq process at the LHC. Clearly, presence
of EFT effects which is already constrained by the existing data can be found in tHq process even for $\rm \mathcal{L}=300~fb^{-1}$ option at the LHC RUN3 experiment. 

\section{Summary}
We present a detailed study of 
new physics effects in the single top quark production associated with the SM Higgs
boson (tHq) within the SMEFT framework where extra dimension-6 terms are added with
the SM lagrangian. As a first step, we isolate a subset of SMEFT operators 
relevant to the tHq process using symmetry arguments. The processes involving a top
quark and having common interactions with the tHq are also identified, namely
$\rm t\bar tH, tH, t\bar tW, tZ$ and tj. The total cross-section or 
differential cross-section, signal strength, etc., are measured 
for each of these processes at the LHC.
The deviation of these latest measurements from the SM prediction are expected to
impose constraints on the WCs corresponding to those SMEFT operators involved in the tHq process. Of course, the sensitivity of the SMEFT operators with the given set of processes determines how strongly those WCs can be constrained.
The FIM is constructed in order to have a better  
quantitative understanding about the level of sensitivity of each of the 
considered processes on various operators interesting to us. It is also a very useful tool to understand our results qualitatively. 
The methodology of the evaluation of the tree level cross-sections(LO) 
for each of the process including SMEFT operators 
are discussed. The compact analytical form of the total cross-section or any observable
can be expressed in terms of a polynomial. The numerical values of the 
coefficients of the respective SMEFT terms are obtained by fitting the variation of cross-section
with the corresponding WC.   
The constrained ranges of chosen SMEFT operators 
are obtained following the technique of $\chi^2$-minimization 
inputting the measured values along with the uncertainties and 
the corresponding theoretical predictions of several observables. 
Eventually, the best fit values of each of these six WCs are presented.  

The tHq process is found to be somewhat special as it incorporates unitarity violating 
scattering sub-amplitude, namely $\rm bW\to tH$, resulting in an energy growth, which is 
reflected in various kinematic distributions as the excess of events at the tail.
This excess of events can be treated as the signature of new physics 
arising due to the EFT contribution. We perform a dedicated simulation of tHq 
process at the LHC energy, $\sqrt{s}=13$ TeV, for both the hadronic and leptonic channels according
to the decay of top quark. The simulation is performed categorizing the events into the boosted and non boosted regions, depending on the $\rm p_T$ of the Higgs boson. For the hadronic final state, the signal is characterized by one reconstructed Higgs-jet, one reconstructed Top-jet and at least one extra jet; whereas, for the leptonic final state, an isolated lepton and an extra b-jet is selected instead of a Top-jet. While in the non-boosted scenario, defined by $\rm p_T(H)<300~GeV$, EFT effects in different distributions are found to be small compared to the SM distributions. In contrast, these effects turn out to be huge in the boosted region ($\rm p_T(H)>300~GeV$).

Having these interesting observations, we concentrate on the boosted region to find the discovery potential of the EFT effects. A detailed simulation of backgrounds are performed to find the signal significance. It is to be noted that the backgrounds are also simulated including the contributions of SMEFT operators.
Systematically selection cuts are applied to suppress backgrounds without costing signal significantly.
Instead of counting total number of excess events in the signal region after background rejection, we present it in bins of a few interesting observables. For hadronic case, the significances are presented in the $\rm p_T$ bins of Higgs and Top jets whereas for leptonic case, $\rm p_T$ of Higgs jet and lepton are considered. The $\rm p_T$-binned analysis turns out to be very effective to single out excesses at the tails of the distributions. It is seen that, a reasonable signal significance $\sim 5 \sigma$ can be achieved in most of the higher $\rm p_T$-bins at the high luminosity option $\rm \mathcal{L}=300~fb^{-1}$, presumably which will further increase at HL-LHC option $\rm \mathcal{L}=3000~fb^{-1}$. In conclusion, our analysis shows that, the scope of the tHq process in boosted scenario is very promising in order to observe discoverable excess due to EFT effects. A more systematic simulation using the p-p collision data can reveal the existence of SMEFT effects, if exist, in the tHq process. We would like to draw the attention of our experimental colleagues in this regard. However, the non-observation of any effect can put severe constraints on the relevant sensitive operators considered in this study.

\acknowledgments{Authors are thankful to Prof. Fabio Maltoni and Prof. Carlos Wagner for valuable suggestions and also to Dr. Suman Chatterjee, Dr. Saikat Karmakar, Dr. Ken Mimasu and Kelci Mohrman for useful discussions. AR also acknowledges  High-Performance Computing facility at TIFR for making high volume of computations possible and `Infosys-TIFR Leading Edge Travel Grant' for providing support to present this study at the ICHEP, 6-13th July 2022, held at Bologna, Italy.}

\appendix
\section{Signal strength in terms of WCs including EFT}

The signal strength ($\mu$) of any given process can be split into the production and decay part, and the decay part is expressed in terms of the decay widths of the final state particles. As an example, we have considered here the decay of the Higgs boson. But, in general, the decay of any particle, such as the top quark or Z boson, etc. can be considered.
\br
\mu^{th}_{xx} =\frac{\sigma^{EFT}(pp\to h)}{\sigma^{SM}(pp\to h)}\times \left(\frac{\Gamma_x^{EFT}(c_i)}{\Gamma_x^{SM}}\right)\times\left(\frac{\Gamma_{tot}^{EFT}(c_i)}{\Gamma_{tot}^{SM}}\right)^{-1}\nonumber \\
=\frac{\sigma^{EFT}}{\sigma^{SM}}\times \left(\frac{\Gamma_x^{EFT}(c_i)}{\Gamma_x^{SM}}\right)\times\left(\frac{\sum_{y}\Gamma_{y}^{EFT}(c_i)}{\sum_{z}\Gamma_{z}^{SM}}\right)^{-1}
\er
where x, y, z are possible final states.

In the framework of EFT expansion, using Eq.~\ref{eq:amp}, we can write both the production cross-section and decay width in the following form,
\br
\sigma^{EFT}(\vec{C})&=&\sigma^{SM}+\sum_{i=1}^{n}a_i^\sigma c_i+\sum_{i\leq j}b_{ij}^\sigma c_i c_j,\label{eq:sigmah}\\
\Gamma_x^{EFT}(\vec{C})&=&\Gamma_x^{SM}+\sum_{i=1}^{n}a_i^\Gamma c_i+\sum_{i\leq j}b_{ij}^\Gamma c_i c_j,
\label{eq:gammah}
\er
where, $a_i^\sigma$ and $b_{ij}^\sigma$ are polynomial coefficients for production, while $a_i^\Gamma$ and $b_{ij}^\Gamma$ are those for decay width. Using Eq.~\ref{eq:mu},\ref{eq:sigmah} and \ref{eq:gammah} one can write,
\br
\mu^{th}_{xx}(\vec{C})=\left(\frac{\sigma^{SM}+\sum_{i=1}^{N}a_i^\sigma c_i+\sum_{j\leq k}b_{jk}^\sigma c_j c_k}{\sigma^{SM}}\right)\times \left(\frac{\Gamma_x^{SM}+\sum_{l=1}^{N}a_l^{\Gamma_x} c_l+\sum_{m\leq n}b_{mn}^{\Gamma_x} c_m c_n}{\Gamma_x^{SM}}\right)\nonumber\\
\times\left(\frac{\sum_{y}\left(\Gamma_y^{SM}+\sum_{r=1}^{N}a_r^{\Gamma_y} c_r+\sum_{s\leq t}b_{st}^{\Gamma_y} c_s c_t\right)}{\sum_{y}\Gamma_z^{SM} }\right)\nonumber\\
=\left(1+\sum_{i}\left(\frac{a_i^{\sigma}}{\sigma^{SM}}\right)c_i+\sum_{j\leq k}\left(\frac{b_{jk}^{\sigma}}{\sigma^{SM}}\right)c_jc_k\right)\times \left(1+\sum_{l}\left(\frac{a_l^{\Gamma_x}}{\Gamma_x^{SM}}\right)c_l+\sum_{m\leq n}\left(\frac{b_{mn}^{\Gamma_x}}{\Gamma_x^{SM}}\right)c_mc_n\right)\nonumber\\
\times \left(\sum_{y}\left[\frac{\Gamma_y^{SM}}{\Gamma_{tot}^{SM}}+\sum_{r}\left(\frac{a_r^{\Gamma_y}}{\Gamma_{tot}^{SM}}\right)c_r+\sum_{s\leq t}\left(\frac{b_{st}^{\Gamma_y}}{\Gamma_{tot}^{SM}}\right)c_sc_t\right]\right)^{-1}\nonumber\\
=\left(1+\sum_{i}\left(\frac{a_i^{\sigma}}{\sigma^{SM}}\right)c_i+\sum_{j\leq k}\left(\frac{b_{jk}^{\sigma}}{\sigma^{SM}}\right)c_jc_k\right)\times \left(1+\sum_{l}\left(\frac{a_l^{\Gamma_x}}{\Gamma_x^{SM}}\right)c_l+\sum_{m\leq n}\left(\frac{b_{mn}^{\Gamma_x}}{\Gamma_x^{SM}}\right)c_mc_n\right)\nonumber\\
\times \left(1-\sum_{y}\left[\sum_{r}\left(\frac{a_r^{\Gamma_y}}{\Gamma_{tot}^{SM}}\right)c_r+\sum_{s\leq t}\left(\frac{b_{st}^{\Gamma_y}}{\Gamma_{tot}^{SM}}\right)c_sc_t\right]\right)\nonumber
\er
\br
=1+\sum_{i}c_i\left[\frac{a_i^{\sigma}}{\sigma^{SM}}+\frac{a_i^{\Gamma_x}}{\Gamma_x^{SM}}-\sum_{y}\left(\frac{a_i^{\Gamma_{y}}}{\Gamma_{tot}^{SM}}\right)\right]+\sum_{j\leq k}c_jc_k\left[\left(\frac{a_j^{\sigma}a_k^{\Gamma_{x}}}{\sigma^{SM}\Gamma_{x}^{SM}}\right)
-\sum_{y}\left(\frac{a_j^{\sigma}a_k^{\Gamma_{y}}}{\sigma^{SM}\Gamma_{tot}^{SM}}\right)\right.\nonumber\\
\left.-\sum_{y}\left(\frac{a_j^{\Gamma_x}a_k^{\Gamma_{y}}}{\Gamma_x^{SM}\Gamma_{tot}^{SM}}\right)+\frac{b_{jk}^{\sigma}}{\sigma^{SM}}+\frac{b_{jk}^{\Gamma_x}}{\Gamma_{x}^{SM}}-\sum_{y}\left(\frac{b_{jk}^{\Gamma_y}}{\Gamma_{tot}^{SM}}\right)\right]\nonumber\\
=1+\sum_{i}A_ic_i+\sum_{j\leq k}B_{jk} c_ic_j.~~~~~~~~~~~~~~~~~~~~~~~~~~~~~~~~~~~~~~~~~~~~~~~~~~~~~~~~~~~~~~~~~~~~~~~~~~~~~~~~~
\er
Where,
\br
A_i&=&\frac{a_i^{\sigma}}{\sigma^{SM}}+\frac{a_i^{\Gamma_x}}{\Gamma_x^{SM}}-\sum_{y}\left(\frac{a_i^{\Gamma_{y}}}{\Gamma_{tot}^{SM}}\right)\\
B_{jk}&=&\left(\frac{a_j^{\sigma}a_k^{\Gamma_{x}}}{\sigma^{SM}\Gamma_{x}^{SM}}\right)
-\sum_{y}\left(\frac{a_j^{\sigma}a_k^{\Gamma_{y}}}{\sigma^{SM}\Gamma_{tot}^{SM}}\right)
-\sum_{y}\left(\frac{a_j^{\Gamma_x}a_k^{\Gamma_{y}}}{\Gamma_x^{SM}\Gamma_{tot}^{SM}}\right)+\frac{b_{jk}^{\sigma}}{\sigma^{SM}}+\frac{b_{jk}^{\Gamma_x}}{\Gamma_{x}^{SM}}-\sum_{y}\left(\frac{b_{jk}^{\Gamma_y}}{\Gamma_{tot}^{SM}}\right)~~~~~~~
\er

\bibliographystyle{JHEP.bst}
\bibliography{eft.bib}
\end{document}